\pdfoutput=1
\documentclass[journal,10pt]{IEEEtran}

\usepackage{empheq,amssymb}
\usepackage{braket}
\usepackage{dsfont,mathrsfs}
\usepackage{enumerate}
\usepackage{bm}
\usepackage{graphicx}
\usepackage{amsmath}

\newtheorem{definition} {Definition}

\newtheorem{theorem}    {Theorem}
\newtheorem{lemma}      {Lemma}

\def\DEFINED    {\coloneqq}

\newcommand\EXP[1]{\mathop{\kern0pt \mathds E}{\Set{#1}}}
\newcommand\PR [1]{\mathop{\kern0pt \Pr}{\Set{#1}}}

\newcommand{\leftexp}[2]%
  {\mathop{}%
   \mathopen{\vphantom{#2}}^{#1}%
   \kern-\scriptspace%
   #2}

\makeatletter
\newcommand\SEQ{\@ifstar\SEQB\SEQA}
\makeatother

\newcommand\SEQA[2][T]{\{#2_t$, $t=1,\dots,#1\}}
\newcommand\SEQB[1]{\{#1_1$, $t=1,\dots\}}

\begin{document}
\title {On the Structure of Real-Time Encoders and Decoders in a Multi-Terminal Communication System}

\author{Ashutosh~Nayyar,~\IEEEmembership{Student Member,~IEEE,}
        and~Demosthenis~Teneketzis,~\IEEEmembership{Fellow,~IEEE}
\thanks{A. Nayyar is with the Department
of Electrical Engineering and Computer Science, University of Michigan, Ann Arbor,
MI, 48109 USA e-mail: (anayyar@umich.edu).}
\thanks{D. Teneketzis is with the Department
of Electrical Engineering and Computer Science, University of Michigan, Ann Arbor,
MI, 48109 USA e-mail: (teneket@eecs.umich.edu).}}

\maketitle

\begin{abstract}
 A real-time communication system with two encoders communicating with a single receiver over separate noisy channels is considered. The two encoders make distinct partial observations of a Markov source. Each encoder must encode its observations into a sequence of discrete symbols. The symbols are transmitted over noisy channels to a finite memory receiver that attempts to reconstruct some function of the state of the Markov source. Encoding and decoding must be done in real-time, that is, the distortion measure does not tolerate delays. Under the assumption that the encoders' observations are conditionally independent Markov chains given an unobserved time-invariant random variable, results on the structure of optimal real-time encoders and the receiver are obtained. It is  shown that there exist finite-dimensional sufficient statistics for the encoders. The problem with noiseless channels and perfect memory at the receiver is then considered. A new methodology to find the structure of optimal real-time encoders is employed. A sufficient statistic with a time-invariant domain is found for this problem. This methodology exploits the presence of common information between the encoders and the receiver when communication is over noiseless channels.
\end{abstract}
\section{Introduction}
  A large variety of decentralized systems require communication between various devices or agents. In general, since such systems may have multiple senders and receivers of information, the models of point-to-point communication are not sufficient. Typically in decentralized systems, the purpose of communication is to achieve a higher system objective.  Examples include networked control systems where the overall objective of communication between various sensors and controllers is to control the plant in order to achieve a performance objective, or sensor networks where the goal of communication between sensors and a fusion center may be to quickly estimate a physical variable or to track in real-time the evolution of a physical phenomenon. In such systems, agents (sensors, controllers etc.) have to make decisions that affect the overall system performance based only on information they currently have gathered from the environment or from other agents through the underlying communication system. The communication problem therefore should not only address what information can be made available to each agent but also when is this information available. Thus, the overall system objectives may impose constraints on the time delay associated with communication.
\par
 In the presence of strict delay constraints on information transmission, the communication problem becomes drastically different from the classical information-theoretic formulations. Information theory deals with encoding and decoding of long sequences which inevitably results in large undesirable delays. For systems with fixed (and typically small) delay requirements, the ideas of asymptotic typicality can not be used. Moreover, information-theoretic bounds on the trade-off between delay and reliability are only asymptotically tight and are of limited value for short sequences (\cite{Gallager}). Therefore, we believe that the development of a real-time communication theory can significantly contribute  to our fundamental understanding of the operation of decentralized systems.
\par
In this paper we address some issues in multi-terminal communication systems under the real-time constraint. Specifically, we look at problems with multiple senders/encoders communicating with a single receiver. We analyze systems with two encoders as in Figure 1, although our results generalize to $n$ encoders $(n>2)$ and a single receiver. The two encoders make distinct partial observations of a discrete-time Markov source. Each encoder must encode in real-time its observations into a sequence of discrete variables that are transmitted over separate noisy channels to a common receiver. The receiver must estimate, in real-time, a given function of the state of the Markov source. The main feature of this multi-terminal problem that distinguishes it from a point to point communication problem is the presence of coupling between the encoders (that is, each encoder must take into account what other encoder is doing). This coupling arises because of the following reasons : 1) The encoders' observations are correlated with each other. 2) The encoding problems are further coupled because the receiver wants to minimize a non-separable distortion metric. That is, the distortion metric cannot be simplified into two separate functions each one of which depends only on one encoder's observations. The nature of optimal strategies strongly depends on the nature and extent of the coupling between the encoders. 
\par     
    Our model therefore involves real-time distributed coding of a pair of correlated observations that are to be transmitted over noisy channels. Information-theoretic results on asymptotically achievable rate-regions have been known for some distributed coding problems.  The first available results on distributed coding of correlated memoryless sources appear in \cite{Slepian} and \cite{Slepian2}. Multiple access channels with arbitrarily correlated sources were considered in \cite{Cover}. In \cite{Flynn}, the encoders make noisy observations of an i.i.d source. The authors in \cite{Flynn} characterize the achievable rates and distortions, and propose two specific distributed source coding techniques. Constructive methods for distributed source coding were presented in \cite{Effros}, \cite{AlJabri} and \cite{Pradhan-Ramchandran}. In particular, \cite{Effros} address lossless and nearly lossless source coding for the multiple access system, and \cite{AlJabri} addresses zero-error distributed source coding. The CEO problem, where a number of encoders make conditionally independent observations of an i.i.d source, was presented in \cite{CEO}. The case where the number of encoders tends to infinity was investigated there. The quadratic Gaussian case of the CEO problem has been investigated in \cite{GaussianCEO}, \cite{Oohama1} and \cite{Draper}. Bounds on the achievable rate-regions for finitely many encoders were found in \cite{CEO2}. A lossy extension of the Slepian-Wolf problem was analyzed in \cite{Zamir}. Multi-terminal source coding for memoryless Gaussian sources was considered in \cite{Oohama2}.
\par
In \cite{Dinesh-Pradhan},\cite{Pinsker},\cite{Korner},\cite{Han},\cite{Ahlswede} and \cite{Csiszar}, distributed source coding problems with the objective of reconstructing a function of the source are investigated. In \cite{Dinesh-Pradhan}, the authors consider distributed source coding of a pair of correlated Gaussian sources. The objective is to reconstruct a linear combination of the two sources. The authors discover an inner bound on the optimal rate-distortion region and provide a coding scheme that achieves a portion of this inner bound.  The problem of distributed source coding to reconstruct a function of the
sources losslessly was considered in \cite{Pinsker}. An inner bound was obtained for the
performance limit which was shown to be optimal if the sources are conditionally
independent given the function. The case of lossless reconstruction of the modulo-2 sum of two correlated binary
sources was considered in \cite{Korner}. These results were extended in \cite{Csiszar} (see Problem 23 on page 400) and \cite{Han}. An improved  inner bound for the problem in \cite{Korner} was provided in \cite{Ahlswede}. 
 \par
 The real-time constraint of our problem differentiates it from the information-theoretic results mentioned above. Real-time communication problems for point-to-point systems have been studied using a decision-theoretic/stochastic control perspective. In general, two types of results have been obtained for point to point systems. One type of results establish qualitative properties of optimal encoding and decoding strategies. The central idea here has been to consider the encoders and the decoders as control agents/decision-makers in a team trying to optimize a common objective of minimizing a distortion metric between the source and its estimates at the receiver. Such sequential dynamic teams - where the agents sequentially make multiple decisions in time and may influence each other's information - involve the solution of non-convex functional optimization to find the best strategies for the agents (\cite{Ho1980},\cite{Witsenhausen68}). However, if the strategies of all but one of the agents are fixed, the resulting problem of optimizing a single agent's strategy can, in many cases, be posed in the framework of Markov decision theory. This approach can explain some of the structural results obtained in \cite{Witsenhausen78},\cite{Teneketzis06},\cite{Walrand83},\cite{Borkar}, \cite{Yuksel}. Another class of results establish a decomposition of  the problem of choosing a sequence of globally optimal encoding and decoding functions. In the resulting decomposition, at each step, the optimization is over one encoding and decoding functions instead of a sequence of functions. This optimization, however, must be repeated for all realizations of an information state that captures the effect of past encoding/decoding functions (\cite{Walrand83},\cite{Borkar},\cite{Mahajan08},\cite{Mahajan-p2p}).
\par
Point to point communication problems with the real-time or finite delay constraint were also investigated from an information-theoretic point of view. We refer the reader to \cite{Teneketzis06} for a survey of the information-theoretic approaches for point-to-point systems with the real-time or finite delay constraint.
\par
  Inspired by the decision-theoretic approach to real-time point-to-point systems, we look at our problem from a decentralized stochastic control/team-theoretic perspective with the encoders and the receiver as our control agents/decision makers.  \emph{We are primarily interested in discovering the structure of optimal real-time encoding and decoding functions.} In other words, given all the observations available to an agent (i.e, an encoder or the receiver), what is a sufficient statistic to decide its action (i.e, the symbol to be transmitted in case of the encoders and the best estimate in case of the receiver)?. The structure of optimal real-time encoding and decoding strategies provides insights into their essential complexity (for example, the memory requirements at the encoders and the receiver for finite and infinite time horizon communication problems) as well as the effect of the coupling between the encoders mentioned earlier. 
  \par
 A universal approach  for discovering the structure of optimal real-time encoding/decoding strategies in a multi-terminal system with any general form of correlation between the encoders' observations has so far remained elusive. In this paper, we restrict ourselves to a simple model for the encoders' observations. For such a model (described in Section \ref{sec:P1}), we obtain results on the structure of optimal real-time encoding strategies when the receiver is assumed to a have a finite memory.  Our results reveal that for any time horizon, however large (or even infinite), there exists a finite dimensional sufficient statistic for the encoders. This implies that an encoder with a memory that can store a fixed finite number of real-numbers can perform as well as encoders with arbitrarily large memories. Subsequently, we consider communication with noiseless channels and remove the assumption of having limited receiver memory. For this problem, the approach in Section \ref{sec:P1} results in sufficient statistics for the encoders that belong to spaces which keep increasing with time. This is undesirable if one wishes to look at problems with large/infinite time-horizons. In order to obtain a sufficient statistic with time-invariant domain, we invent a new methodology for decentralized decision problems. This methodology highlights the importance of common information/ common knowledge (in the sense of \cite{Aumann76}), in determining structural properties of decision makers in a team. In general, the resulting sufficient statistic belongs to an infinite dimensional space. However, we present special cases where a finite dimensional representation is possible. Moreover, we believe that the infinite dimensional sufficient statistic may be intelligently approximated to obtain real-time finite-memory encoding strategies whose performance is close to optimal. 
\par
The rest of the paper is organized as follows: In Section \ref{sec:GP} we present a real-time multi-terminal communication system and formulate the optimization problem. In Section \ref{sec:P1} we present our assumptions on the nature of the source and the receiver and obtain structural results for optimal real-time encoding and decoding strategies. In Section \ref{sec:P2} we consider the problem with noiseless channels and perfect receiver memory. We develop a new methodology to find structural results for optimal real-time encoders for this case. We look at some extensions and special cases of our results in Section \ref{sec:Ext}. We conclude in Section \ref{sec:Conc}.
\par
\emph{Notation:}
1. Throughout this paper, subscripts of the form \(1:t\), like \(X_{1:t}\), are used to denote sequences like  \(X_1, X_2,..,X_t\). \\
2. We denote random variables with capital letters \((X)\) and their realization with small letters \((x)\). For random vectors, we add a tilde \(( \hspace{2pt}\tilde{}\hspace{2pt} )\) over the vector to denote its realization.\\
3. For continuous random-variables (or vectors), \(P(X=x)\) refers to \(P(x \leq X < x+dx)\). \\
4. For a set \(\mathcal{A}\), we use \(\Delta(\mathcal{A})\) to denote the space of probability densities (or probability mass functions) on \(\mathcal{A}\).

\section{A Real-Time Multi-terminal Communication Problem}
\label{sec:GP}
  Consider the real-time communication system shown in Figure 1. We have two encoders that partially observe a Markov source and communicate it to a single receiver over separate noisy channels. The receiver may be interested in estimating the state of the Markov source or some function of the state of the source. We wish to find sufficient statistics for the encoders and the receiver and/or qualitative properties for the encoding and decoding functions. 
Below, we elaborate on the model and the optimization problem.
 \begin{figure}[ht]
\begin{center}

\includegraphics[height=4cm,width=8cm]{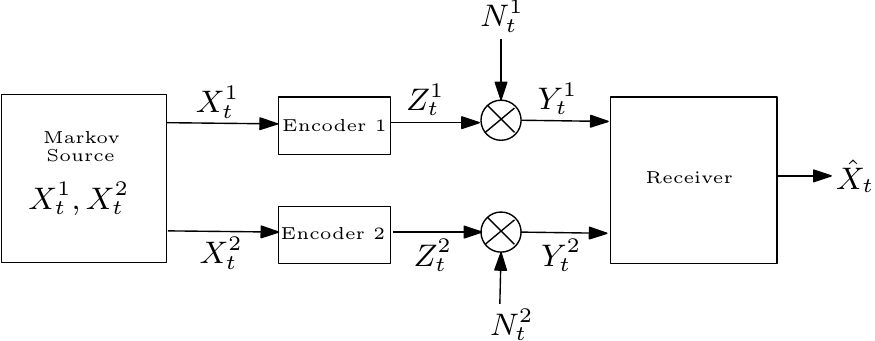}
\caption{A Multi-terminal Communication System}

\end{center} 
\end{figure}  
  \subsection{Problem Formulation}
\emph{1) The Model:} The state of the Markov source at time \(t\) is described as
 \[ X_t = (X^1_t,X^2_t)\]
 where \(X^i_t \in \mathcal{X}^{i}\), \(i=1,2\) and \(\mathcal{X}^{1}, \mathcal{X}^{2}\) are finite spaces. The time-evolution of the source is given by the following equation
 \begin{equation}
  X_{t+1} = F_t(X_t,W_t)
 \end{equation}
 where \(W_t,t=1,2,..\) is a sequence of independent random variables that are independent of the initial state \(X_1\).
 \par
   Two encoders make partial observations of the source. In particular, at time \(t\), encoder 1 observes \(X^1_{t}\) and encoder 2 observes \(X^2_{t}\). The encoders have perfect memory, that is, they remember all their past observations and actions. At each time \(t\), encoder 1 sends a symbol \(Z^1_{t}\) belonging to a finite alphabet \(\mathcal{Z}^1\) to the receiver. The encoders operate in real-time, that is, each encoder can select the symbol to be sent at time \(t\), based only on the information available to it till that time. That is, the encoding rule at time \(t\) must be of the form:
         \begin{equation} \label{eq:encoder1}
         Z^1_{t}=f^1_{t}(X^1_{1:t},Z^1_{1:t-1})
         \end{equation}
         where \(X^1_{1:t}\) represents the sequence \(X^1_{1},X^1_{2},\ldots,X^1_{t}\) and \(Z^1_{1:t-1}\) represents the sequence \(Z^1_{1},Z^1_{2},\ldots,Z^1_{t-1}\). In general, one can allow randomized encoding rules instead of deterministic encoding functions. That is, for each realization of its observations till time \(t\), encoder 1 selects a probability distribution on \(\mathcal{Z}^1\) and then transmits a random symbol generated according to the selected distribution.  We will show later that, under our assumptions on the model, such randomized encoding rules cannot provide any performance gain and we can restrict our attention to deterministic encoding functions.\\
         Encoder~2 operates in a similar fashion as encoder 1. Thus, encoding rules of encoder~2 are functions of the form:
           \begin{equation}\label{eq:encoder2}
         Z^2_{t}=f^2_{t}(X^2_{1:t},Z^2_{1:t-1})
         \end{equation}
         where \(Z^2_{t}\) belongs to finite alphabet \(\mathcal{Z}^2\). 
         \par
         The symbols \(Z^1_{t}\) and \(Z^2_{t}\) are transmitted over separate noisy channels to a single receiver. The channel noises at time \(t\) are mutually independent random variables \(N^1_{t}\) and \(N^2_{t}\) belonging to finite alphabets \(\mathcal{N}^1\) and \(\mathcal{N}^2\) respectively. The noise variables \((N^1_{1},N^2_{1},N^1_2,N^2_2,\ldots,N^1_{t},N^2_t,\ldots)\) form a collection of independent random variables that are independent of the source process \(X_t, t=1,2,...\).
         \par
         The receiver receives \(Y^1_{t}\) and \(Y^2_{t}\) which belong to finite alphabets \(\mathcal{Y}^1\) and \(\mathcal{Y}^2\) respectively. The received symbols are noisy versions of the transmitted symbols according to known channel functions \(h^1_{t}\) and \(h^2_{t}\), that is,
         \begin{equation}
         Y^i_{t}=h^i_{t}(Z^i_{t},N^i_{t})
         \end{equation}
         for \(i=1,2\).
        
         At each time \(t\), the receiver produces an estimate of the source \(\hat{X}_{t}\) based on the symbols received till time \(t\), i.e.,
         \begin{equation}\label{eq:receiverPG}
         \hat{X}_{t}=g_{t}(Y^1_{1:t},Y^2_{1:t})
         \end{equation}
          A non-negative distortion function \(\rho_t(X_t,\hat{X_t})\) measures the instantaneous distortion between the source and the estimate at time \(t\). (Note that the distortion function may take into account that the receiver only needs to estimate a function of \(X^1_t\) and \(X^2_t\))

\par

\emph{ 2) The Optimization Problem P}:
Given the source and noise statistics, the encoding alphabets, the channel functions \(h^1_{t},h^2_{t}\), the distortion functions \(\rho_{t}\) and a time horizon T, the objective is to find globally optimal encoding and decoding functions \(f^1_{1:T},f^2_{1:T},g_{1:T}\) so as to minimize
          \begin{equation} \label{eq:Objective0.1}
                    J(f^1_{1:T},f^2_{1:T},g_{1:T})= \EXP{\sum\limits_{t=1}^{T} \rho_{t}(X_{t},\hat{X_{t}})}
          \end{equation}
          where the expectation in (\ref{eq:Objective0.1}) is over the joint distribution of \(X_{1:T}\) and \(\hat{X}_{1:T}\) which is determined by the given source and noise statistics and the choice of encoding and decoding functions \(f^1_{1:T},f^2_{1:T},g_{1:T}\). \\
          We refer to the collection of functions \(f^i_{1:T}\) as encoder~\(i\)'s strategy (\(i=1,2\)). The collection of functions  \(g_{1:T}\) is the decoding strategy.\\
          Remarks: 1. Since we consider only finite alphabets for the source, the encoded symbols, the channel noise, the received symbols  and a finite time horizon, the number of possible choices of encoding and decoding functions is finite. Therefore, an optimal choice of strategies \((\tilde{f}^{1}_{1:T},\tilde{f}^{2}_{1:T},\tilde{g}_{1:T})\) always exists. \\
          2. A brute force search method to find the optimal can always be used in principle. It is clear however that even for small time-horizons, the number of possible choices would be large enough to make such a search inefficient. Moreover, such a scheme would not be able to identify any characteristics of optimal encoding and decoding functions.     \par 
             The encoding functions and the decoding functions in equations (\ref{eq:encoder1}), (\ref{eq:encoder2}) and (\ref{eq:receiverPG}) require the encoders and the receiver to store entire sequences of their past observations and actions. For large time-horizons storing all past data becomes prohibitive. Therefore, one must decide what part of the information contained in these arbitrarily large sequences is sufficient for decision-making at the encoders and the receiver. In particular, we are  interested in addressing the following questions:
             \begin{enumerate}
\item Is there a sufficient statistic for the encoders and the decoder that belongs to a time-invariant space? (Clearly, all the past data available at an agent is a sufficient statistic but it belongs to a space that keeps increasing with time.) If such a sufficient statistic exists, one can potentially look at problems with large (or infinite) time-horizons.
\item Is there a finite-dimensional sufficient statistic for the encoders and the receiver? If such a sufficient statistic exists, then we can replace the requirement of storing arbitrarily long sequences of past observations/messages with storing a fixed finite number of real numbers at the encoders and the receiver.
\end{enumerate} 
\par
        The above communication problem can be viewed as a sequential team problem where the encoders and the receiver are the decision-making agents that are sequentially making decisions  to optimize a common objective. The communication problem is a dynamic team problem since the encoders' decisions influence the information available to the receiver. Dynamic team problems are known to be hard. For dynamic teams, a general answer to the questions on the existence of sufficient statistics that either have time-invariant domains or are finite-dimensional is not known. In the next section we will make simplifying assumptions on the nature of the source and the receiver and present sufficient statistics for the encoders.

\section{Problem P1}\label{sec:P1}
      We consider the optimization problem (Problem P) formulated in the previous section under the following assumptions on the source and the receiver. \\
\emph{1. Assumption A1 on the Source:} We assume that the time-evolution of the source can be described by the following model:
\begin{subequations}
\begin{equation}
    X^1_{t+1} = F^1_{t}(X^1_t,A,W^1_t)
\end{equation}
\begin{equation}
X^2_{t+1} = F^2_{t}(X^2_t,A,W^2_t)
\end{equation}
\end{subequations}
where \(A\) is a random-variable taking values in  the finite set \(\mathcal{A}\) and  \(W^1_t, t=1,2,...\) and \(W^2_t, t=1,2...\) are two independent noise processes (that is, sequences of independent random variables) that are independent of the initial state \((X^1_1,X^2_1\) and \(A)\) as well. Thus, the transition probabilities satisfy:
        \begin{align}
      &P(X^1_{t+1},X^2_{t+1}|X^1_t,X^2_t,A) \nonumber \\
        = &P(X^1_{t+1}|X^1_t,A).P(X^2_{t+1}|X^2_t,A) \label{eq:source2}
         \end{align}
         The initial state of the Markov source has known statistics that satisfy the following equation :
         \begin{align}
        P(X^1_1,X^2_1,A) = &P(X^1_1,X^2_1|A).P(A) \nonumber \\
        = &P(X^1_1|A).P(X^2_1|A).P(A) \label{eq:source1}
        \end{align}
        Thus, \(A\) is a time-invariant random variable that couples the evolution of \(X^1_t\) and \(X^2_t\).  Note that conditioned on \(A\), \(X^1_t\) and \(X^2_t\) form two conditionally independent Markov chains. We define \begin{equation}
        X_t := (X^1_t,X^2_t,A)
        \end{equation}
         which belongs to the space \(\mathcal{X} := \mathcal{X}^1 \times \mathcal{X}^2 \times \mathcal{A}\).\\
        The encoders' model is same as before. Thus encoder 1 observes \(X^1_t\) and encoder 2 observes \(X^2_t\). Note that the random variable \(A\) is not observed by any encoder. The encoders have perfect memories and the encoding functions are given by equations (\ref{eq:encoder1}) and (\ref{eq:encoder2}). 
        \par
        \emph{2. Assumption A2 on the Receiver:} We have a finite memory receiver that maintains a separate memory for symbols received from each channel. This memory is updated as follows:
         \begin{subequations}\label{eq:memory update}
         \begin{equation}
         M^i_{1}=l^i_1(Y^i_{1}), i=1,2
         \end{equation}
         \begin{equation}
         M^i_{t}=l^i_t(M^i_{t-1},Y^i_{t}), i=1,2
         \end{equation}
         \end{subequations}
         where \(M^i_{t}\) belongs to finite alphabet \(\mathcal{M}^i\), \(i=1,2\) and \(l^i_t\) are the memory update functions at time \(t\) for \(i=1,2\). For notational convenience, we define \(M^i_0 := 0\) for \(i=1,2\). The receiver produces an estimate of the source \(\hat{X}_{t}\) based on its memory contents at time \(t-1\) and the symbols received at time \(t\), that is,
         \begin{equation}\label{eq:receiverP1}
         \hat{X}_{t}=g_{t}(Y^1_{t},Y^2_{t},M^1_{t-1},M^2_{t-1})
         \end{equation}
         We now formulate the following problem.
         \par
\emph{Problem P1:} With assumptions A1 and A2 as above, and given source and noise statistics, the encoding alphabets, the channel functions \(h^1_{t},h^2_{t}\), the distortion functions \(\rho_{t}\) and a time horizon T, the objective is to find globally optimal encoding, decoding and memory update functions \(f^1_{1:T},f^2_{1:T},g_{1:T},l^1_{1:T},l^2_{1:T}\) so as to minimize
          \begin{equation} \label{eq:Objective1}
                    J(f^1_{1:T},f^2_{1:T},g_{1:T},l^1_{1:T},l^2_{1:T})= \EXP{\sum\limits_{t=1}^{T} \rho_{t}(X_{t},\hat{X_{t}})}
          \end{equation}
          where the expectation in (\ref{eq:Objective1}) is over the joint distribution of \(X_{1:T}\) and \(\hat{X}_{1:T}\) which is determined by the given source and noise statistics and the choice of encoding, decoding  and memory update functions \(f^1_{1:T},f^2_{1:T},g_{1:T},l^1_{1:T},l^2_{1:T}\).
          \begin{figure}[ht]
\begin{center}

\includegraphics[height=4cm,width=8cm]{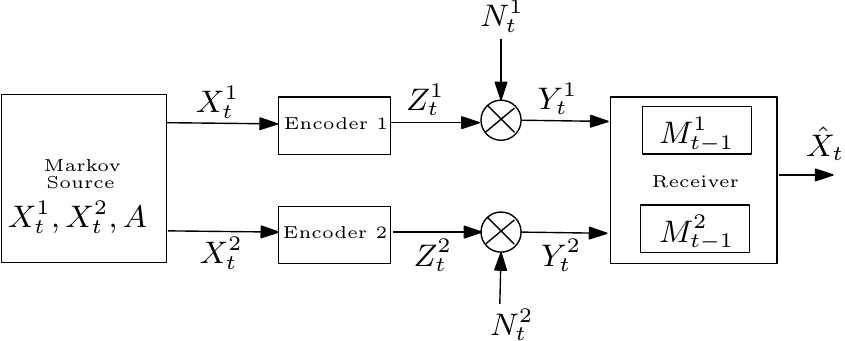}
\caption{Problem P1}

\end{center} 
\end{figure}  
\subsection{Features of the Model}
We discuss situations that give rise to models similar to that of Problem P1.
\par
 1. A Sensor Network:  Consider a sensor network where the sensors' observations are influenced by a slowly varying global parameter and varying local phenomena. Our model is an approximation of this situation where \(A\) models the global parameter that is constant over the time-horizon \(T\) and \(X^i_t\) are the local factors at the location of the \(i^{th}\) sensor at time \(t\). A finite memory assumption on the receiver may be justified in situations where the receiver is itself a node in the network  and is coordinating the individual sensors. We will show that this assumption implies that the sensors (encoders in our model) themselves can operate on finite-dimensional sufficient statistics without losing any optimality with respect to sensors with perfect (infinite) memory.
\par
2. Decentralized Detection/Estimation Problem: Consider the following scenario of a decentralized detection problem; Sensors make noisy observations \(X^i_t\) on the state \(A\) of environment. Sensors must encode their information in real-time and send it to a fusion center. Assuming that sensor noises are independent, we have that, conditioned on \(A\), the sensor observations are independent. (Typically, the observations are also assumed to be i.i.d in time conditioned on the state of the environment, but we allow them to be Markov.) Thus, the encoding rule for the \(i^{th}\) sensor must be of the form:
     \[Z^i_t=f^i_t(X^i_{1:t},Z^i_{1:t-1})\]
 Consider the case where \(Z^i_t\) can either be ``blank'' or a value from the set \(\mathcal{A}\). Each sensor is restricted to send only one non-blank message, and within a fixed time-horizon each sensor must send its final non-blank message. When a sensor sends a non-blank message \(Z^i_t\), the fusion center receives a noisy version \(Y^i_t\) of this message.  As long as the fusion center does not receive final (non-blank) messages from all sensors, its decision is \(\hat{X}_t =\) ``no decision'' and the system incurs a constant penalty \(c\) (for delaying the final decision on \(A\)). If all sensors have sent a non-blank message, the fusion center produces an estimate \(\hat{X}_t \in \mathcal{A}\)  as its final estimate on \(A\) and incurs a distortion cost \(\rho(A,\hat{X}_t)\). Thus, we can view the receiver as maintaining a separated memory for messages from each sensor which is initialized to ``blank'' and  updated as follows:
     \begin{align} \label{eq:detection} 
        M^i_t = \left \{ \begin{array}{ll}
               Y^i_t & \mbox{if $M^i_{t-1}$ was ``blank''} \\
               M^1_{t-1} & \mbox{otherwise}
               \end{array}
               \right. \end{align}
         The receiver's decision is \( \hat{X}_t =\) ``no decision'', if \(Y^i_t = M^i_{t-1}= \) ``blank'' for some sensor \(i\), else the receiver uses a function \(g_t\) to find an estimate
         \begin{equation}
         \hat{X}_t = g_{t}(Y^1_{t},Y^2_{t},M^1_{t-1},M^2_{t-1})
         \end{equation}
         The above detection problem therefore is a special case of our model with fixed memory update rules from (\ref{eq:detection}). 
\par
 Clearly, our model also includes the case when the encoders' observations are independent Markov chains (not just conditionally independent). In this case, the coupling between encoders is only due to the fact the receiver may be interested in estimating some function of the state of the two Markov chains and not their respective individual states.          

\subsection{Structure Result for Encoding Functions}
We define the following probability mass functions (pmf) for encoder~\(i\), \((i=1,2)\):
\begin{definition}
For \(t=1,2,\ldots,T\) and \( a \in \mathcal{A}\),
\[ b^i_t(a) \DEFINED P(A=a|X^i_{1:t})  \]
\end{definition}
\begin{definition}
For \(t=2,3,\ldots,T\) and \( m \in \mathcal{M}^i\),
\[ \mu^i_{t}(m) \DEFINED P(M^i_{t-1}=m|Z^i_{1:t-1}, l^i_{1:t-1}) \]
where \(l^i_{1:t-1}\) in the conditioning indicate that \(\mu^i_t\) is defined for a fixed choice of the memory update rules \(l^i_{1:t-1}\).
For notational convenience, we also define for each \(m \in \mathcal{M}^i, i=1,2\),
\[ \mu^i_{1}(m) \DEFINED 0\]
\end{definition}
\par
\begin{theorem}
  There exist globally optimal encoding rules of the  form  :
   \begin{equation}
   Z^i_{t}=f^i_{t}(X^{i}_{t},b^i_{t},\mu^i_{t}) \label{eq:theorem one}
   \end{equation}
   where \(f^i_t\) are deterministic functions for \(t=1,2,\ldots,T\) and \(i=1,2\).
\end{theorem}
\par
\emph{Discussion:} In contrast to equation (\ref{eq:encoder1}), Theorem 1 says that an optimal encoder~1 only needs to use the current observation \(X^1_t\) and the probability mass functions \(b^1_t, \mu^1_t\) that act as a compressed representation of the past observations \(X^1_{1:t-1}\) and \(Z^1_{1:t-1}\). These pmfs represent the encoder~1's belief on \(A\) and \(M^1_{t-1}\).
\par
To obtain the result of Theorem 1 for the encoder 1, we fix arbitrary encoding rules for the encoder 2 of the form in (\ref{eq:encoder2}), arbitrary memory update rules of the form in (\ref{eq:memory update}) and arbitrary decoding rules of the form in (\ref{eq:receiverP1}). Given these functions, we consider the problem of selecting optimal encoding rules for encoder 1. We identify a  structural property of the optimal encoding rules of encoder 1  that is independent of the arbitrary choice of strategies for encoder 2 and the receiver. We conclude that the identified structure of optimal rules of encoder 1 must also be true when encoder 2 and the receiver are using the globally optimal strategies. Hence, the identified structure is true for globally optimal encoding rules of encoder 1. We now present this argument in detail.
\par
Consider arbitrary (but fixed) encoding rules for encoder 2 of the of the form in (\ref{eq:encoder2}), arbitrary memory update rules for the receiver of the form in (\ref{eq:memory update}) and arbitrary decoding rules of the form in (\ref{eq:receiverP1}). We will prove Theorem 1 using the following lemmas.
\begin{lemma}
    The belief of the first encoder about the random variable \(A\) can be updated as follows:
    \begin{equation}
    b^1_{t}=\alpha^1_{t}(b^1_{t-1},X^1_{t},X^1_{t-1})
    \end{equation}
    where \(\alpha^1_{t},t=2,3,\ldots,T\) are deterministic functions.
\end{lemma}
\begin{proof}
See Appendix A.
\end{proof}
\begin{lemma}
    The belief of the first encoder about the receiver memory \(M^1_{t-1}\) can be updated as follows:
    \begin{equation}
    \mu^1_{t}=\beta^1_{t}(\mu^1_{t-1},Z^1_{t-1})
    \end{equation}
 where \(\beta^1_{t},t=2,3,\ldots,T\) are deterministic functions.
\end{lemma}
\begin{proof}
See Appendix B.
\end{proof}
\par
\vspace{8pt}
    Define the following random variables:
    \begin{align}
     R^1_{t} &\DEFINED (X^1_{t},b^1_{t},\mu^1_{t}),
    \end{align}
     for \(t=1,2,\ldots,T\).\\
Observe that \(R^1_{t}\) is a function of encoder 1's observations till time \(t\), that is, \(X^1_{1:t},Z^1_{1:t-1}\). Moreover, any encoding rule of the form in (\ref{eq:encoder1}) can also be written as
\[ Z^1_{t}=f^1_{t}(R^1_{1:t},Z^1_{1:t-1}) \]
\begin{lemma}
    \(R^1_{t}, t=1,2,...,T\) is a perfectly observed controlled Markov process for encoder 1 with \(Z^1_{t}\) as the control action at time \(t\).
\end{lemma}
\begin{proof}
     Since \(R^1_{t}\) is a function of encoder 1's observations till time \(t\), that is, \(X^1_{1:t},Z^1_{1:t-1}\), it is perfectly observed at encoder~1. \\
     Let \(x^1_{1:t},z^1_{1:t-1}\) be a realization of the encoder 1's observations \(X^1_{1:t},Z^1_{1:t-1}\). Similarly, let \(r^1_{t}\) be a realization of \(R^1_{t}\) and \(\tilde{b}^1_{t}\) and \(\tilde{\mu}^1_{t}\) be realizations of \(b^1_{t}\) and \(\mu^1_{t}\) respectively. Then,
    \begin{eqnarray}
     \lefteqn{P(R^1_{t+1}=(x^1_{t+1},\tilde{b}^1_{t+1},\tilde{\mu}^1_{t+1})|r^1_{1:t},z^1_{1:t})} \nonumber \\
    &=&P(x^1_{t+1},\tilde{b}^1_{t+1},\tilde{\mu}^1_{t+1}|x^1_{1:t},\tilde{b}^1_{1:t},\tilde{\mu}^1_{1:t},z^1_{1:t})\nonumber \\
     &=&P(\tilde{b}^1_{t+1},\tilde{\mu}^1_{t+1}|x^1_{t+1}, x^1_{1:t},\tilde{b}^1_{1:t},\tilde{\mu}^1_{1:t},z^1_{1:t})      \nonumber \\
     & & \times P(x^{1}_{t+1}|x^1_{1:t},\tilde{b}^1_{1:t},\tilde{\mu}^1_{1:t},z^1_{1:t}) \label{eq:proof1.1}\\
     &=&P(\tilde{b}^1_{t+1},\tilde{\mu}^1_{t+1}|x^1_{t+1},x^1_{t},\tilde{b}^1_{t},\tilde{\mu}^1_{t},z^1_{t})     \nonumber \\
     & & \times P(x^{1}_{t+1}|x^1_{1:t},\tilde{b}^1_{1:t},\tilde{\mu}^1_{1:t},z^1_{1:t})   \label{eq:proof1.2}
     \end{eqnarray}
     where the first term in (\ref{eq:proof1.2}) is true because of Lemma 1 and Lemma 2.
     Consider the second term in (\ref{eq:proof1.2}). It can be expressed as follows:
     \begin{align}
     \lefteqn{P(x^{1}_{t+1}|x^1_{1:t},\tilde{b}^1_{1:t},\tilde{\mu}^1_{1:t},z^1_{1:t})} \nonumber \\
     = \sum\limits_{a \in\mathcal{A}} &P(x^{1}_{t+1},A=a|x^1_{1:t},\tilde{b}^1_{1:t},\tilde{\mu}^1_{1:t},z^1_{1:t})\\
     = \sum\limits_{a \in\mathcal{A}} &P(x^1_{t+1}|A=a,x^1_{1:t},\tilde{b}^1_{1:t},\tilde{\mu}^1_{1:t},z^1_{1:t}) \nonumber\\
      \times &P(A=a|x^1_{1:t},\tilde{b}^1_{1:t},\tilde{\mu}^1_{1:t},z^1_{1:t})\\
     = \sum\limits_{a \in \mathcal{A}} &P(x^1_{t+1}|A=a,x^1_{t}).\tilde{b}^1_{t}(a) \label{eq:proof1.3}
     \end{align}
     where the first term in (\ref{eq:proof1.3}) is true because of the Markov property of \(X^1_{t}\) when conditioned on \(A\). Therefore, substituting (\ref{eq:proof1.3}) in (\ref{eq:proof1.2}), we get
     \begin{align}
     &P(R^1_{t+1}=(x^1_{t+1},\tilde{b}^1_{t+1},\tilde{\mu}^1_{t+1})|x^1_{1:t},\tilde{b}^1_{1:t},\tilde{\mu}^1_{1:t},z^1_{1:t})  \nonumber \\
     = &P(\tilde{b}^1_{t+1},\tilde{\mu}^1_{t+1}|x^1_{t+1},x^1_{t},\tilde{b}^1_{t},\tilde{\mu}^1_{t},z^1_{t})     \nonumber \\
      &\times \sum\limits_{a \in \mathcal{A}} [P(x^1_{t+1}|A=a,x^1_{t}) \times \tilde{b}^1_{t}(a)] \label{eq:proof1.3a}  
     \end{align}
     The right hand side of (\ref{eq:proof1.3a}) depends only on \(x^1_{t}\),\(\tilde{b}^1_{t},\tilde{\mu}^1_{t}\) and \(z^1_{t}\) from the entire collection of conditioning variables in the left hand side of (\ref{eq:proof1.3a}). Hence,
     \begin{align}
      P(R^1_{t+1}|r^1_{1:t},z^1_{1:t}) = &P(R^1_{t+1}|x^1_{1:t},\tilde{b}^1_{1:t},\tilde{\mu}^1_{1:t},z^1_{1:t}) \nonumber \\
      = &P(R^1_{t+1}|x^1_{t},\tilde{b}^1_{t},\tilde{\mu}^1_{t},z^1_{t}) \nonumber \\
      = &P(R^1_{t+1}|r^1_{t},z^1_{t})
     \end{align}
     This establishes the Lemma.
\end{proof}

\begin{lemma}
    The expected instantaneous distortion cost for encoder 1 can be expressed as :
    \begin{equation}
    \EXP{\rho_{t}(X_{t},\hat{X}_{t})|X^1_{1:t},Z^1_{1:t}} = \hat{\rho}_{t}(R^1_{t},Z^1_{t})
    \end{equation}
    where \(\hat{\rho}_t, t=1,2,\ldots,T\) are deterministic functions.
\end{lemma}
\begin{proof} For any realization \(x^1_{1:t},z^1_{1:t}\) of \(X^1_{1:t},Z^1_{1:t}\), we have
    \begin{align}
     &\EXP{\rho_{t}(X_{t},\hat{X}_{t})|x^1_{1:t},z^1_{1:t}} \nonumber \\
    = &\EXP{\rho_{t}(x^1_{t},X^2_{t},A,g_{t}(Y^1_{t},Y^2_{t},M^1_{t-1},M^2_{t-1})|x^1_{1:t},z^1_{1:t})} \label{eq:proof1.4}
    \end{align}
    The expectation in (\ref{eq:proof1.4}) depends on \(x^1_{t}\) (appearing in the argument of \(\rho_t\)) and the conditional probability: \(P(X^2_{t},A,Y^1_{t},Y^2_{t},M^1_{t-1},M^2_{t-1}|x^1_{1:t},z^1_{1:t})\). We can evaluate this conditional probability as follows:
    \begin{align}
        &P(X^2_{t}=x^2_{t},A=a,Y^1_{t}=y^1_{t},Y^2_{t}=y^2_{t}, \nonumber \\
        &M^1_{t-1}=m^1_{t-1},M^2_{t-1}=m^2_{t-1}|x^1_{1:t},z^1_{1:t}) \label{eq:P1lemma4.2} \\
        = &P(X^2_{t}=x^2_{t},Y^2_{t}=y^2_{t},M^2_{t-1}=m^2_{t-1}|A=a,Y^1_{t}=y^1_{t}, \nonumber
        \\
        &M^1_{t-1}=m^1_{t-1},x^1_{1:t},z^1_{1:t}) \times \nonumber \\
          &P(Y^1_{t}|A=a,M^1_{t-1}=m^1_{t-1},x^1_{1:t},z^1_{1:t}) \times \nonumber \\
          &P(M^1_{t-1}=m^1_{t-1}|A=a,x^1_{1:t},z^1_{1:t}) \times \nonumber \\
          &P(A=a|x^1_{1:t},z^1_{1:t}) \\
       =  &P(X^2_{t}=x^2_{t},Y^2_{t}=y^2_{t},M^2_{t-1}=m^2_{t-1}|A=a) \times \nonumber \\ &P(Y^1_{t}=y^1_{t}|z^1_{t})\times P(M^1_{t-1}=m^1_{t-1}|z^1_{1:t})\times P(A=a|x^1_{1:t}) \label{eq:P1lemma4.1} \\
       =  &P(X^2_{t}=x^2_{t},Y^2_{t}=y^2_{t},M^2_{t-1}=m^2_{t-1}|A=a) \times \nonumber \\
       &P(Y^1_{t}=y^1_{t}|z^1_{t})\times \tilde{\mu}^1_{t}(m^1_{t-1})\times \tilde{b}^1_{t}(a) \label{eq:proof1.5}
    \end{align}
    In the first term of (\ref{eq:P1lemma4.1}), we used the fact that conditioned on \(A\), the observations of encoder 2 and received messages from the second channel are independent of the observations of encoder 1 and the messages received from the first channel. We used the fact that the noise variables \(N^1_t\) are i.i.d and independent of the source in the second and third term of (\ref{eq:P1lemma4.1}). Thus, the conditional probability in (\ref{eq:P1lemma4.2}) depends only on \(z^1_{t},\tilde{\mu}^1_{t}\) and \(\tilde{b}^1_{t}\). Therefore, the expectation in (\ref{eq:proof1.4}) is a function of \(x^1_{t},z^1_{t},\tilde{\mu}^1_{t},\tilde{b}^1_{t}\). That is,
    \begin{align}
    &\EXP{\rho_{t}(X_{t},\hat{X}_{t})|x^1_{1:t},z^1_{1:t}} =\hat{\rho}_{t}(x^1_{t},z^1_{t},\tilde{\mu}^1_{t},\tilde{b}^1_{t})\\
    & = \hat{\rho}_{t}(r^1_{t},z^1_{t})
    \end{align}
\end{proof}
\begin{proof}[Proof of Theorem 1] From Lemma 3 and Lemma 4, we conclude that the optimization problem for encoder 1, when the strategies of encoder 2 and the receiver have been fixed, is equivalent to controlling the transition probabilities of the controlled Markov chain \(R^1_{t}\) through the choice of the control actions \(Z^1_t\) (where \(Z^1_{t}\) can be any function of \(R^1_{1:t}\) and \(Z^1_{1:t-1}\)) in order to minimize \( \sum_{t=1}^{T}\EXP{\hat{\rho}_{t}(R^1_{t},Z^1_{t})}\). It is a well-known result of Markov decision theory (\cite{Kumarvaraiya}, Chapter 6) that there is an optimal control law of the form:
\[Z^1_{t}=f^1_{t}(R^1_{t})\]
or equivalently,
\[ Z^1_{t}=f^1_{t}(X^{1}_{t},b^1_{t},\mu^1_{t})\]
Moreover, it also follows from Markov decision theory that allowing randomized control policies for encoder 1 cannot provide any performance gain.
Since the above structure of the optimal choice of encoder~1's strategy is true for any arbitrary choice of encoder~2's and the receiver's strategies, we conclude that the above structure of optimal encoder~1 is true when the encoder~2 and the receiver are using their globally optimal choices as well. Therefore, the above structure is true for globally optimal strategy of encoder~1 as well.  This completes the proof of Theorem~1. Structural result for encoder~2 follows from the same arguments simply by interchanging the roles of encoder~1 and encoder~2.
\end{proof}

\subsection{Structural result for Decoding Functions}
We now present the structure of an optimal decoding strategy. Consider fixed encoding rules of the form in (\ref{eq:encoder1}) and (\ref{eq:encoder2}) and fixed memory update rules of the form in (\ref{eq:memory update}). We define the following probability mass function for the receiver :
\begin{definition}
For \(x \in \mathcal{X}\) and \(t=1,2, \ldots ,T\),
\[\psi_{t}(x) \DEFINED P(X_{t}=x|Y^1_{t},Y^2_{t},M^1_{t-1},M^2_{t-1},f^1_{1:t},f^2_{1:t},l^1_{1:t},l^2_{1:t})\]
\end{definition}
where the functions \( f^1_{1:t},f^2_{1:t},l^1_{1:t},l^2_{1:t}\) in the conditioning indicate that \(\psi_{t}\) is defined for a fixed choice of encoding and memory update strategies.
\par
Let \(\Delta( \mathcal{X})\) denote the set of probability mass functions on the finite set \(\mathcal{X}\). We define the following functions on \(\Delta(\mathcal{X})\).
\begin{definition}
For any \(\psi \in \Delta( \mathcal{X} )\) and \(t=1,2,\ldots ,T\),
\[\tau_{t}(\psi) = \underset{s \in \mathcal{X}}{\operatorname{argmin}} \sum\limits_{x\in \mathcal{X}}\psi(x)\rho_{t}(x,s)\]
\end{definition}

With the above definitions, we can present the result on the structure of a globally optimal decoding rule.\\
\par
\begin{theorem}
For any fixed encoding rules of the form in (\ref{eq:encoder1}) and (\ref{eq:encoder2}) and memory update rules of the form in (\ref{eq:memory update}), there is an optimal decoding rule of the form
\begin{equation} \label{eq:theorem2}
\hat{X}_{t} = \tau_{t}(\psi_{t})
\end{equation}
where the belief \(\psi_{t}\) is formed using  the fixed encoding and memory update rules. In particular, equation (\ref{eq:theorem2}) is true for a globally optimal receiver, when the fixed encoding rules and memory update rules are the globally optimal rules.
\end{theorem}
\begin{proof}
  In order to minimize the expected total accumulated distortion, the receiver must minimize the expected distortion at each time \(t\). Clearly, the definitions of the function \(\tau_t\) and the belief \(\psi_t\) imply that \(\tau_t(\psi_t)\) achieves the minimum expected distortion at time \(t\) (see \cite{Teneketzis06}). 
\end{proof}
\subsection{Discussion of the Result}
  Theorem 1 identifies sufficient statistics for the encoders. Instead of storing all past observations and transmitted messages, each encoder may store only the probability mass functions (pmf) on the finite sets \(\mathcal{A}\) and \(\mathcal{M}^i\) generated from past observations and transmitted messages. Thus we have finite-dimensional sufficient statistics for the encoders that belong to time-invariant spaces (the space of pmfs on \(\mathcal{A}\) and \(\mathcal{M}^i\)). Clearly, this amounts to storing a fixed number of real-numbers in the memory of each encoder instead of arbitrarily large sequences of past observations and past transmitted symbols.  However, the encoders now have to incur an additional computational burden involved in updating their beliefs on \(A\) and the receiver memory.  
 \par
 We would like to emphasize that the presence of a finite dimensional sufficient statistic that belong to time-invariant spaces is strongly dependent on the nature of the source and the receiver. Indeed, without the conditionally independent nature of the encoders' observations or the separated finite memories at the receiver, we have not been able to identify a sufficient statistic whose domain does not keep increasing with time. For example, if the finite memory receiver maintained a coupled memory which is updated as:
 \[ M_t =l_t(M_{t-1},Y^1_t,Y^2_t) \]
 then one may conjecture that the encoder could use a belief on \(M_{t-1}\) as a sufficient representation of past transmitted symbols, analogous to \(\mu^1_t\) in Theorem 1. However, such a statistic cannot be updated without remembering all past data, that is, an update equation analogous to Lemma 2 for \(\mu^1_t\) does not hold. This implies that the Markov decision-theoretic arguments of Theorem 1 do not work for this case.\\
  In the case when encoders' observations have a more general correlation structure, a finite dimensional statistic like \(b^1_t\) that compresses all the past observations seems unlikely. It appears that in the absence of the assumptions mentioned above, the optimal encoders should remember all their past information.  
  \par
  If the receiver has perfect memory, that is, it remembers all past messages received, \((M^i_{t-1} = Y^i_{1:t-1}, i=1,2)\), Theorem 1 implies \(\mu^i_t=P(Y^i_{1:t-1}|Z^i_{1:t-1})\) as a part of the sufficient statistic for encoder \(i\). Thus, Theorem 1 says that each encoder needs to store beliefs on the increasing space of all past observations at the receiver. This sufficient statistic does not belong to a time-invariant space. In the next section, we will consider this problem with noiseless channels and show that for noiseless channels there is in fact a sufficient statistic that belongs to a time-invariant space. However, this sufficient statistic is no longer finite dimensional and for implementation purposes, one would have to come up with approximate representations of it.

\section{Problem P2}
\label{sec:P2}

 		We now look at the Problem P1 with noiseless channels. Firstly, we assume the same model for the nature of the source and the separated memories at the receiver as in Problem P1. The result of Theorem 1 holds with the belief on \(M^i_{t-1}\) replaced by the true value of \(M^i_{t-1}\). The presence of noiseless channels implies that encoder \(i\) and the receiver have some common information. That is, at time \(t\) they both know the state of \(M^i_{t-1}\). The presence of common information among agents of a team allows for new ways of optimizing the team objective (\cite{Nayyar-Allerton08}). In this section, we will show that the presence of common information allows us to explore the case when the receiver may have perfect memory. We will present a new methodology that exploits the presence of common information between the encoder and the receiver to find sufficient statistics for the encoders that belong to time-invariant spaces (spaces that do not keep growing with time). 

\subsection{Problem Formulation}
\begin{enumerate}
\item \emph{The Model:} We consider the same model as in P1 with following two modifications:
\begin{enumerate}[i.]
\item The channels are noiseless; thus the received symbol \(Y^i_{t}\) is same as the transmitted symbol \(Z^i_{t}\), for \(i=1,2\) and \(t=1,2, \ldots, T\).
\item The receiver has perfect memory, that is, it remembers all the past received symbols. Thus, \(M^i_{t-1}=Z^i_{1:t-1}\), for \(i=1,2\) and \(t=2,3, \ldots, T\). (See Fig. 3)
\end{enumerate}
\item \emph{The Optimization Problem, P2}:
Given the source statistics, the encoding alphabets, the time horizon T,  the distortion functions \(\rho_{t}\), the objective is to find globally optimal encoding and decoding functions \(f^1_{1:T},f^2_{1:T},g_{1:T}\) so as to minimize
          \begin{equation} \label{eq:Objective2}
                    J(f^1_{1:T},f^2_{1:T},g_{1:T})=E[\sum\limits_{t=1}^{T} \rho_{t}(X_{t},\hat{X_{t}})]
          \end{equation}
          where the expectation in (\ref{eq:Objective2}) is over the joint distribution of \(X_{1:T}\) and \(\hat{X}_{1:T}\) which is determined by the given source statistics and the choice of encoding and decoding functions \(f^1_{1:T},f^2_{1:T},g_{1:T}\).
\end{enumerate}
\subsection{Structure of the Receiver}
Clearly, problem P2 is a special case of problem P1. The decoder structure of P1 can now be restated for P2 as follows:
 For fixed encoding rules of the form in (\ref{eq:encoder1}) and (\ref{eq:encoder2}),  we can define the receiver's belief on the source as:\\
\[\psi_{t}(x) \DEFINED P(X_{t}=x|Z^1_{1:t},Z^2_{1:t},f^1_{1:t},f^2_{1:t})\]
for \(x \in \mathcal{X}\) and \(t=1,2, \ldots ,T\).\\

\begin{figure}
\begin{center}
\includegraphics[height=2cm,width=7cm]{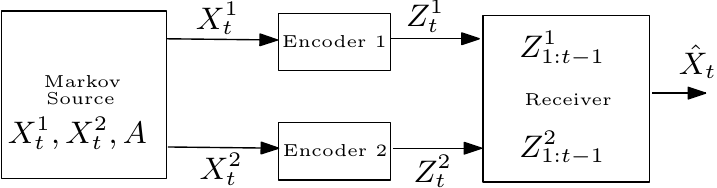}
\caption{Problem P2}
\end{center} 
\end{figure}  
\begin{theorem}
For any fixed encoding rules of the form in (\ref{eq:encoder1}) and (\ref{eq:encoder2}), there is an optimal decoding rule of the form
\begin{equation} \label{eq:theorem3}
\hat{X}_{t} = \tau_{t}(\psi_{t})
\end{equation}
where the belief \(\psi_{t}\) is formed using  the fixed encoding rules and \(\tau_t\) is as defined in Definition 4. In particular, equation (\ref{eq:theorem3}) is true for a globally optimal receiver, when the fixed encoding rules are globally optimal rules.
\end{theorem}
\subsection{Structural Result for Encoding Functions}

  For a fixed realization of \(Z^i_{1:t-1}\), encoder~\(i\)'s belief on the receiver memory \(M^i_{t-1}\) is simply:
 \begin{align}
 \tilde{\mu}^i_{t}(m) = P(M^i_{t-1}=m|z^i_{1:t-1}) = \left \{ \begin{array}{ll}
                                                               1 & \mbox{if $m=z^1_{1:t-1}$} \\
                                                               0 & \mbox{otherwise}
                                                               \end{array} 
                                                               \right.
 \end{align}
Therefore, using Theorem 1, we conclude that
   there is a globally optimal encoder of the form:
\[Z^i_{t}=f^i_{t}(X^{i}_{t},b^i_{t},\mu^i_{t})\] for \(t=1,2,\ldots,T\) and \(i=1,2\).\\
Or equivalently,
\begin{equation}
Z^i_{t}=f^i_{t}(X^{i}_{t},b^i_{t},Z^i_{1:t-1}) \label{eq:P2encoder1}
\end{equation}

Observe that the domain of the encoding functions in (\ref{eq:P2encoder1}) keeps increasing with time since it includes all past transmitted symbols \(Z^1_{1:t-1}\). We would like to find a sufficient statistic that belongs to a time-invariant space. Such a statistic would allow us to address problems with large (or infinite) time horizons.
\par
      For that matter, let us first review the approach used for obtaining the first structural result for the encoders (Theorem 1). We fixed the strategy of encoder~2 and the receiver to any arbitrary choice and looked at the optimization problem P1 from encoder 1's perspective. Essentially, we addressed the following question: if encoder 2 and the receiver have fixed their strategies, how can we characterize the best strategy of encoder 1 in response to the other agents' fixed strategies? In other words, with \(f^2_{1:T}\) and \(g_{1:T}\) as fixed, what kind of strategies of encoder 1 \((f^1_{1:T})\) minimize the objective in equation (\ref{eq:Objective1})? This approach allowed us to formulate a Markov decision problem for encoder 1. The Markov decision problem gave us a sufficient statistic for encoder 1 that holds for any choice of strategies of encoder 2 and the receiver and this led to the result of Theorem 1. In problem P2, such an approach gives the result of equation (\ref{eq:P2encoder1}) - which implies a sufficient statistic whose domain keeps increasing with time. 
\par
     To proceed further, we need to adopt a different approach. As before, we will make an arbitrary choice of encoder~2's strategy of the form in (\ref{eq:encoder2}). Given this fixed encoder 2, we will now ask, what are the jointly optimal strategies for encoder~1 and the receiver? That is, assuming \(f^2_{1:T}\) is fixed, what choice of \(f^1_{1:T}\) and \(g_{1:T}\) together minimize the objective in equation (\ref{eq:Objective2})? From our previous structural results, we know that we can restrict to encoding rules \(f^1_{1:T}\) of the form in (\ref{eq:P2encoder1}) and decoding rules from Theorem 3 without any loss of optimality. We thus have the following problem:\\
     \emph{\underline{Problem P2'}:} In Problem P2, with encoder~2's strategy fixed to an arbitrary choice \(f'^{2}_{1:T}\), find the jointly optimal strategies of encoder~1 of the form in (\ref{eq:P2encoder1}) and of the receiver in Theorem 3 to minimize
     \[  J(f^1_{1:T},f'^{2}_{1:T},g_{1:T})=E[\sum\limits_{t=1}^{T} \rho_{t}(X_{t},\hat{X_{t}})]\]     
     \par
     
     Problem P2' is in some sense a real-time point-to-point communication problem with side information at the receiver. This is now a decentralized team problem with the first encoder and the receiver as the two agents. Note that encoder~1 influences the decisions at the receiver not only by the symbols \(Z^1_t\) it sends but by the entire encoding functions it employs (since the receiver's belief \(\psi_t\) depends on the choice of encoding functions \(f^1_{1:t}\)). A general way to solve such dynamic team problems is to search through the space of all strategies to identify the best choice. For our problem (and for many team problems), this is not a useful approach for two reasons: 1)~Complexity - the space of all strategies is clearly too large even for small time horizons, thus making a brute force search prohibitive. 2)~More importantly, such a method does not reveal any characteristic of the optimal strategies and does not lead to the identification of a sufficient statistic. We will therefore adopt a different philosophy to address our problem. 
     \par
     Our approach is to first consider a modified version of problem P2'. We will construct this modified problem in such a way so as to ensure that:\\
     (a) The new problem is a single agent problem instead of a team problem. Single agent centralized problems (in certain cases) can be studied through the framework of Markov decision theory and dynamic programming. \\
     (b) The new problem is \emph{equivalent} to the original team problem. We will show that the conclusions from the modified problem remain true for the  problem  P2' as well. \\
     We proceed as follows:\\
\emph{Step 1:} We introduce a centralized stochastic control problem from the point of view of a fictitious agent who knows the ``common information'' between encoder~1 and the receiver. \\
\emph{Step 2:} We argue that the centralized problem of Step 1 is equivalent to  the original decentralized team problem.
\emph{Step 3:} We solve  the centralized stochastic control problem by identifying an information state and employing dynamic programming arguments. The solution of this problem will reveal a sufficient statistic with a time-invariant domain for encoder~1.

Below, we elaborate on these steps.
\par
\textbf{Step 1}: We observe that the first encoder and the receiver have some common information. At time \(t\), they both know \(Z^1_{1:t-1}\). We now  formulate a centralized problem from the perspective of a fictitious agent that knows just the common information \(Z^{1}_{1:t-1}\). We call this fictitious agent the ``coordinator'' (See Fig. 4). 
\begin{figure}
\begin{center}
\includegraphics[height=4cm,width=7cm]{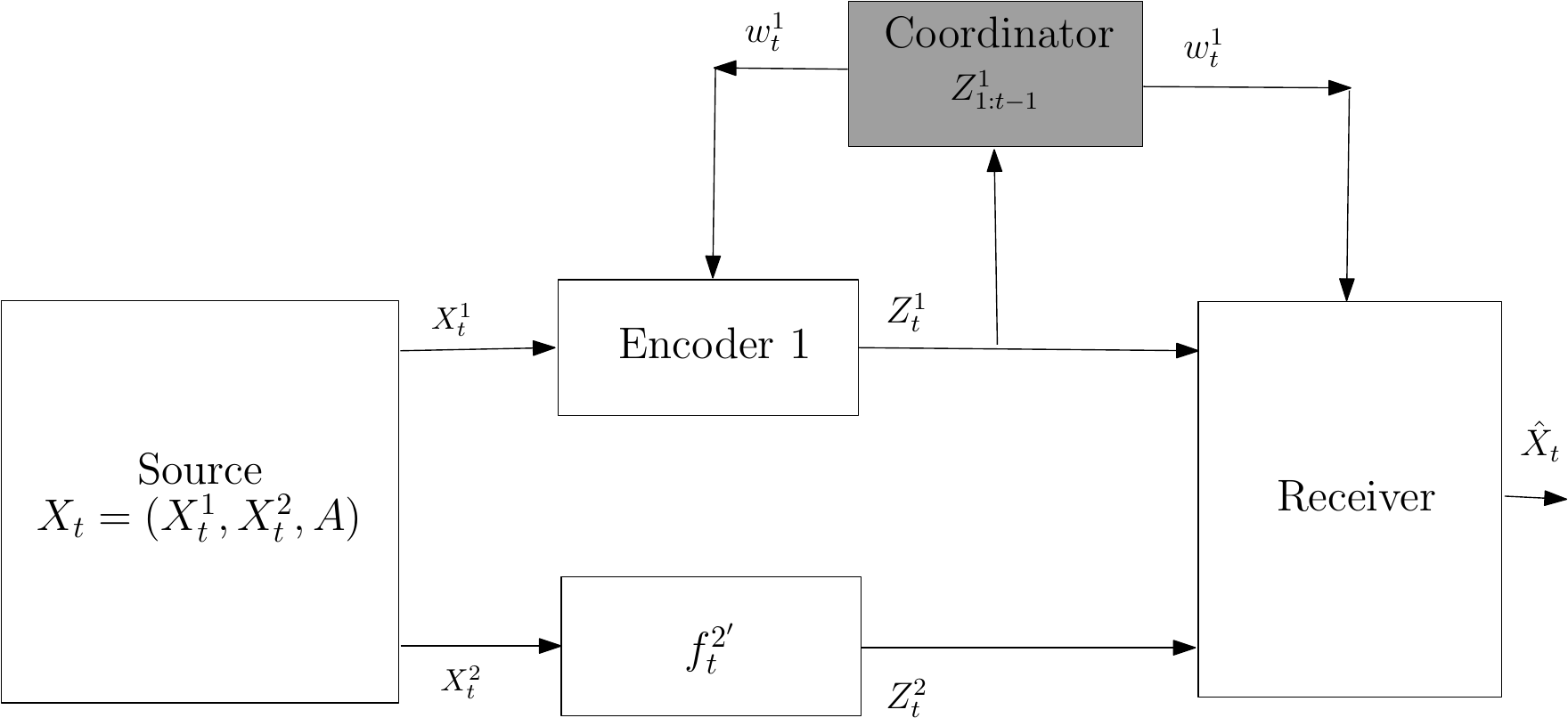}
\caption{Coordinator's Problem P2''}
\end{center} 
\end{figure}
\par
The system operates as follows in this new problem: Based on \(Z^{1}_{1:t-1}\), the coordinator selects a partial-encoding function \[w^1_{t}: \mathcal{X}^{1} \times \Delta(\mathcal{A}) \longrightarrow \mathcal{Z}^{1}\]  
An encoding function of the form in (\ref{eq:P2encoder1}) can be thought of as a collection of mappings from \(\mathcal{X}^{1} \times \Delta(\mathcal{A})\) to
\(\mathcal{Z}^{1}\) - one for each realization of \(Z^1_{1:t-1}\). Clearly, \(w^1_t\) represents one such mapping corresponding to the true realization of \(Z^1_{1:t-1}\) that was observed by the coordinator. (At \(t=1\), since there is no past common information, the partial-encoding rule \(w^1_1\) is simply \(f^1_1\) which is a mapping from \(\mathcal{X}^{1} \times \Delta(\mathcal{A})\) to \(\mathcal{Z}^{1}\).) \\
The coordinator informs the encoder~1 of its choice \(w^1_t\). The encoder~1 then  uses \(w^1_{t}\) on its observations \(X^1_{t}\) and \(b^1_t\) to find the symbol to be transmitted, i.e,
\begin{equation}
Z^1_{t} = w^1_{t}(X^1_{t},b^1_t) \label{eq:P2encoder}
\end{equation}
The coordinator also informs the receiver of the partial-encoding function.
The receiver at each time \(t\), forms its belief on the state of the source based on the received symbols, the partial-encoding functions and the fixed strategy of encoder~2. This belief is  
\[ \psi_t (x) := P(X_{t}=x|z^{1}_{1:t},z^{2}_{1:t},w^{1}_{1:t},f'^{2}_{1:t}) \]
for \(x \in \mathcal{X}\). The receiver's optimal estimate at time \(t\) is then given as:
\begin{equation} \label{eq:coordinator-receiver}
\hat{X}_t  = \underset{s \in \mathcal{X}}{\operatorname{argmin}} \sum\limits_{x\in \mathcal{X}}\psi_{t}(x)\rho_{t}(x,s) \end{equation} 
The coordinator  then observes the symbol \(Z^1_t\) sent  from encoder~1 to the receiver and then selects the partial-encoding function for time \(t+1\) (\(w^1_{t+1}\)). The system continues like this from time \(t=1\) to \(T\).
The objective of the coordinator is to minimize the performance criterion of equation (\ref{eq:Objective2}), that is, to minimize \[E[\sum\limits_{t=1}^{T} \rho_{t}(X_{t},\hat{X_{t}})]\]
We then have the following problem: \\
\emph{\underline{Problem P2''}:} In Problem P2, with encoder~2's strategy fixed to the same choice \(f'^{2}_{1:T}\) as in P2' and with a coordinator between encoder~1 and the receiver as described above, find an optimal selection rule for the coordinator, that is find the mappings \(\Lambda_t, t=1,2,...,T\) that map the coordinator's information to its decision
\[w^1_t = \Lambda_t(Z^1_{1:t-1},w^1_{1:t-1})\]
so as to minimize the total expected distortion over time \(T\). \\
(Note we have included the past decisions (\(w^1_{1:t-1}\)) of the coordinator in the argument of \(\Lambda_t\) since they themselves are functions of  the past observations \(Z^1_{1:t-1}\)). \\
\emph{Remark:} Under a given selection rule for the coordinator, the function \(w^1_t\) is a random variable whose realization depends on the realization of past \(Z^1_{t-1}\) which, in turn, depends on the realization of the source process and the past partial-encoding functions. 
\par
\textbf{Step 2:} We now argue that the original team problem P2' is equivalent to the problem in the presence of the coordinator (Problem P2''). Specifically, we show that any achievable value of the objective (that is, the total expected distortion over time \(T\)) in problem P2' can also be achieved in problem P2'' and vice versa. Consider first any selection rule \(\Lambda_t, t=1,2,...,T\) for the coordinator. While introducing the coordinator in Step 1, we gave it a particular structure- namely, we said that the coordinator only knows the common information between encoder~1 and the receiver. This is crucial because it implies that all information available at the coordinator is in fact available to both encoder~1 and the receiver. Thus, the selection rule \(\Lambda_t\) of the coordinator can be used by both encoder~1 and the receiver to determine the partial-encoding function, \(w^1_t\), to be used at time \(t\) even when the coordinator is not actually present! With encoder~2 fixed as before, the system operation  for the model in Problem P2' can now be described as follows: At each time \(t\), encoder~1, uses the selection rule \(\Lambda_t\) to decide the partial function to be used at time \(t\), it then  uses \(w^1_t\) to evaluate \(Z^1_t\) as follows :
\[ Z^1_t=w^1_t(X^1_t,b^1_t)\]
The receiver uses the same selection rule to find out what \(w^1_t\) is being used by encoder~1. It then uses the received symbols to form its belief on the source and produce an estimate according to equation (\ref{eq:coordinator-receiver}). Therefore, the coordinator can effectively be simulated by encoder~1 and the receiver, and hence any achievable value of the objective in Problem P2'' with the coordinator can be achieved even in the absence of a physical coordinator.
\par
Conversely, in Problem P2' consider any strategy \(f^1_{1:T}\) of encoder~1 and the corresponding optimal receiver given by Theorem 3. Now consider the following selection rule for the coordinator in P2'': At each time \(t\), after having observed \(z^1_{1:t-1}\), the coordinator selects the following partial encoding function.
\[w^1_t(\cdot) = f^1_t(\cdot,z^1_{t-1})\]
Then it is clear that for every realization of the source, encoder~1 in Problem P2'' will produce the same realization of encoded symbols as encoder~1 of Problem P2'. Consequently the  above selection rule of the coordinator  will induce the same joint distribution \(P(X_{1:T},Z^1_{1:T},Z^2_{1:T})\) as the encoding rules \(f^1_{1:T}\) for encoder~1 in problem P2'. Then the receivers in Problem P2' and Problem P2'' will have the same conditional belief \(\psi_t\) and will make the same estimates (given by Theorem 3 and equation (\ref{eq:coordinator-receiver}) respectively). Thus any achievable value of the objective in Problem P2' can also be achieved in Problem P2''. 
\par
  The above equivalence allows us to focus on the coordinator's problem to solve the original problem P2'. We now argue that the coordinator's problem is in fact a single agent centralized problem for which Markov decision-theoretic results can be employed.
  \par
  \textbf{Step 3:}   
 To further describe the coordinator's problem we need the following definition and lemma.
\begin{definition}
For \(t=1,2, \ldots, T\), let \(\xi^1_{t}\) be the coordinator's belief on \(X^1_t,b^1_t\). That is,
\[ \xi^1_t(x^1_t,\tilde{b}^1_t) \DEFINED P(X^1_t=x^1_t,b^1_t=\tilde{b}^1_t|Z^1_{1:t},w^1_{1:t}) \]
for \(x^1_t \in \mathcal{X}^1\) and \(\tilde{b}^1_t \in \Delta(\mathcal{A})\).
\end{definition}
For notational convenience, we define \(\xi^1_0 := 0\).
\begin{lemma}
For a fixed strategy of encoder~2, there is an optimal decoding rule of the form:
\begin{equation}
\hat{X}_t = \tau_t(\psi_t) = \tau_t(\delta_t(\xi^1_t,Z^2_{1:t})) \label{eq:P2lemma5a}
\end{equation}
where \(\delta_{t},t=1,2, \ldots, T\) are fixed transformations that depend only on source statistics and the fixed strategy of encoder~2 and \(\tau_{t}, t=1,2, \ldots, T\) are the decoding functions as defined in \emph{Definition 4}.
\end{lemma}

\begin{proof}
See Appendix C.
\end{proof}
\vspace{8pt}
\par
 From equations (\ref{eq:P2encoder}) and (\ref{eq:P2lemma5a}), it follows that in the coordinator's problem P2'', encoder 1 and the receiver are simply implementors of fixed transformations. They do not make any decisions. Thus, in this formulation, the coordinator is the sole decision maker. We now analyze the centralized problem for the coordinator.
 
\par
 Firstly, observe that at time \(t\), the coordinator knows its observations so far - \(Z^1_{1:t-1}\) and the partial encoding functions it used so far - \( w^1_{1:t-1}\); it then selects an ``action'' \(w^1_t\) and makes the next ``observation'' \(Z^1_t\). In particular, note that the coordinator has perfect recall, that is, it remembers all its past observations and actions-this is a critical characteristic of classical centralized problems for which Markov decision-theoretic results hold.
\par
We can now prove the following lemma :
\begin{lemma}
\begin{enumerate}
\item With a fixed strategy of the second encoder, \(\xi^1_t\) can be updated as follows:
\begin{equation}
\xi^1_t=\gamma^1_t(\xi^1_{t-1},Z^1_t,w^1_t)  \label{eq:P2lemma6a}
\end{equation}
where \(\gamma^1_{t},t=2, \ldots, T\) are fixed transformations that depend only on the source statistics.
\item For a fixed strategy of the second encoder, the expected instantaneous cost from the coordinator's perspective can be written as:
\begin{equation}
\EXP{\rho_{t}(X_{t},\hat{X}_{t})|Z^1_{1:t},w^1_{1:t}} = \overline{\rho}_{t}(\xi^1_t) \label{eq:P2lemma6b}
\end{equation}
for \(t=1,2, \ldots, T\), where \(\overline{\rho}_{t}\) are deterministic functions.
\end{enumerate}
\end{lemma}
\begin{proof}
See Appendix D.
\end{proof}

Based on Lemma 6, we obtain the following result on the coordinator's optimization problem.
\begin{theorem}
For any given selection rule \(\Lambda_t, t=1,2...,T\) for the coordinator, there exists another selection rule \(G_t,t=1,2,...,T\) that selects the partial-encoding function to be used at time \(t\), \((w^1_t)\) based only on \(\xi^1_{t-1}\) and whose performance is no worse than that of \(\Lambda_t, t=1,2,...,T\). Therefore, one can optimally restrict to selection rules for the coordinator of the form:
\begin{equation}
w^1_t = G_t(\xi^1_{t-1})
\end{equation}

\end{theorem}
\begin{proof}
Because of Lemma 6, the optimization problem for the coordinator is to control the evolution of \(\xi^1_t\) (given by (\ref{eq:P2lemma6a})) through its actions \(w^1_t\), when the instantaneous cost depends only on  \(\xi^1_t\). Since \(\xi^1_t\) is known to the coordinator, this problem is similar to the control of a perfectly observed Markov process. This observation essentially implies the result of the theorem, as it follows from Markov decision theory (\cite{Kumarvaraiya}, Chapter 6) that to control a perfectly observed Markov process one can restrict attention to policies that depend only on the current state of the Markov process without any loss of optimality. A more detailed proof using the backward induction method of dynamic programming is given in Appendix E.

\end{proof}
We have therefore identified the structure of the coordinator's selection rule. The coordinator does not need to remember all of its information - \(Z^1_{1:t-1}\) and \(w^1_{1:t-1}\). It can operate optimally by just using \(\xi^1_{t-1}\).
 We can thus conclude the following result.
 \begin{theorem}
 In Problem P2, there is no loss of optimality in considering decoding rules of the form in Theorem 3 with encoders that operate as follows: \\
 For \(i=1,2\), define \(\xi^i_0 := 0\) and for \(t=1,2,...T\), 
       \begin{equation} \label{eq:P2encoder2}
        Z^i_t =f^i_t(X^i_t,b^i_t,\xi^i_{t-1})
        \end{equation}
        and
        \begin{equation} 
        \xi^i_t =\gamma^i_t(\xi^i_{t-1},Z^i_t,f^i_t(\cdot,\xi^i_{t-1}))
        \end{equation}
        where \(\gamma^i_t\) are fixed transformations (Lemma 6).
 \end{theorem}
   \begin{proof}
       The assertion of the the theorem follows from Theorem 4 and the equivalence between problem P2' and P2'' established in Step 2. The coordinator (either real or simulated by encoder~1 and receiver) can select the partial encoding functions by a selection rule of the form:
       \[w^1_t = G_t(\xi^1_{t-1}) \]
      and the encoder~1's symbol to be transmitted at time \(t\) is given as:
      \[Z^1_t = w^1_t(X^1_t,b^1_t)\]
      Thus, \(Z^1_t\) is a function of \(X^1_t,b^1_t\) and \(\xi^1_{t-1}\) that was used to select \(w^1_t\). That is,
      \[ Z^1_t =f^1_t(X^1_t,b^1_t,\xi^1_{t-1}) \]
      where \(w^1_t(\cdot) = f^1_t(\cdot,\xi^1_{t-1})\).
      The coordinator (real or simulated) then updates \(\xi^1_{t-1}\) according to Lemma 6 as:
      \[\xi^1_t=\gamma^1_t(\xi^1_{t-1},Z^1_t,w^1_t)\]
The same argument holds for encoder~2 as well.
   \end{proof}
\subsection{Discussion}
    
		Observe that \(Z^1_{1:t-1}\) appearing in the argument of optimal encoding functions in (\ref{eq:P2encoder1}) have been replaced by \(\xi^1_{t-1}\). By definition, \(\xi^1_{t}\) is a joint belief on \(\mathcal{X}^1\) and \(\Delta(\mathcal{A})\), therefore, \(\xi^1_t\) belongs to a time-invariant space, namely, the space of joint beliefs on \(\mathcal{X}^1\) and \(\Delta(\mathcal{A})\). Thus the domain of the optimal encoding functions in (\ref{eq:P2encoder2}) is time-invariant. 
However, \(\xi^1_t\) above is a joint belief on a finitely-valued random variable (\(X^1_t\)) and a real-valued vector (\(b^1_{t}\)). Thus, we have an infinite-dimensional sufficient statistic for the encoder. Clearly, such a sufficient statistic can not be directly used for implementation. However, it may still be used in identifying good approximations to optimal encoders. 
Below, we present some cases where the above structural result may suggest finite-dimensional representations of the sufficient statistic.

\subsection{Special Cases}
 \subsubsection{A observed at the Encoders}
           Consider the case when the encoder's observations at time \(t=1\) include the realization of the random variable \(A\). Clearly, the encoder's belief on \(A\), (\(b^i_t\)) can be replaced by the true value of \(A\) in Theorem 5. Thus, for problem P2, there is an optimal encoding rule of the form:
          \begin{equation} \label{eq:Ext2}
						Z^1_t=f^1_t(X^1_t,A,P(X^1_{t-1},A|Z^1_{1:t-1},f^1_{1:t-1}))
					\end{equation}
Since \(A\) belongs to a finite set, the domain of the encoding functions in (\ref{eq:Ext2}) consists of the scalars \(X^1_t\) and \(A\) and  a belief on the finite space \(\mathcal{X}^1 \times \mathcal{A}\). Thus when \(A\) is observed at the encoders, we have a finite dimensional sufficient statistic for each encoder. 
\subsubsection{Independent Observations at Encoders}
  Consider the case when the encoders' observations are independent Markov chains. This is essentially the case when \(A\) is constant with probability 1. Then, effectively, all agents know \(A\).  In this case, the result of (\ref{eq:Ext2}) reduces to
\begin{equation} \label{eq:Ext3}
Z^1_t=f^1_t(X^1_t,P(X^1_{t-1}|Z^1_{1:t-1},f^1_{1:t-1}))
\end{equation}
and we have a finite dimensional sufficient statistic for the encoders.
\subsubsection{Binary \(A\)}
    Consider the case when \(A\) can take only two values : 0 or 1. Then the encoder's belief \(b^i_t\) can be described by a single real number,
    \[ b^i_t := P(A=0|X^i_{1:t})\] 
    Thus, \(\xi^1_{t-1}\) in Theorem 4 involves forming a joint belief on a finitely valued \(X^i_{t-1}\) and a real-number \(b^i_{t-1} \in [0,1]\). Although probability distributions on real number can't be stored in the encoder's memory, we can still work with approximate versions of these beliefs. For example, we may decide to store the cumulative distribution function for only certain values in \([0,1]\) as an approximation of the true distribution. This would give a time-invariant finite dimensional approximation of the encoder's information.  Approximate ways of evaluating and updating these approximate beliefs would be required for this scheme to become feasible.



  
\section{Extensions} \label{sec:Ext}
    We apply our results for Problems P1 to P2 to other related problems in this section.

\subsection{Multiple (n) encoders and single receiver problem}
\begin{figure}[ht]
\begin{center}
\includegraphics[height=4cm,width=7cm]{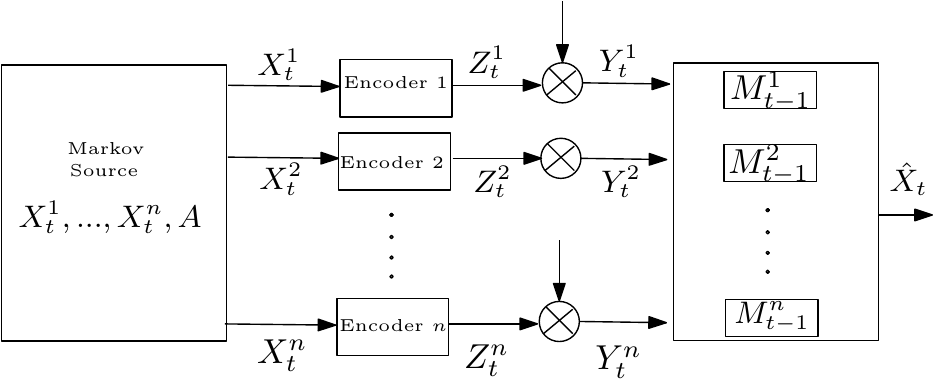}
\end{center}
\caption{Problem with \(n\) encoders} 
\end{figure} 
   Consider the model of Figure 5 where we have \(n\) \((n>2)\) encoders that partially observe a Markov source and encode their observations, in real-time, into sequences of discrete symbols that are transmitted over separate noisy channels (with independent noise) to a common receiver. We make assumptions analogous to assumptions A1 and A2 for Problem P1, that is,
\par
1. Assumption 1: The state of the Markov source is given as : \[X_t := (X^1_t,X^2_t,...,X^n_t,A) \] where \(A\) is a time-invariant random variable and conditioned on \(A\), \(X^1_t,X^2_t,...,X^n_t\) are conditionally independent Markov chains. The \(i^{th}\) encoder observes the process \(X^i_t, t=1,2,...\) and uses encoding functions of the form :\[ Z^i_t =f^i_t(X^i_{1:t},Z^i_{i:t-1}) \] for \(i=1,2,...,n\).
\par
2. Assumption 2: We have a finite memory receiver that maintains a separate memory for symbols received from each channel. This memory is updated as follows:
         \begin{subequations}
         \begin{equation}
         M^i_{1}=l^i_1(Y^i_{1}), i=1,2,...,n
         \end{equation}
         \begin{equation}
         M^i_{t}=l^i_t(M^i_{t-1},Y^i_{t}), i=1,2,...,n
         \end{equation}
         \end{subequations}
         where \(M^i_{t}\) belongs to finite alphabet \(\mathcal{M}^i\),  and \(l^i_t\) are the memory update functions at time \(t\) for \(i=1,2,...,n\). The receiver produces an estimate of the source \(\hat{X}_{t}\) based on its memory contents at time \(t-1\) and the symbols received at time \(t\), that is,
         \begin{equation}
         \hat{X}_{t}=g_{t}(Y^1_{t},Y^2_{t},...,Y^n_t,M^1_{t-1},M^2_{t-1},...,M^n_{t-1})
         \end{equation}
         A non-negative distortion function \(\rho_t(X_t,\hat{X}_t)\) measures the instantaneous distortion at time \(t\).
         We can now formulate the following problem.
         \par
\emph{Problem P3:} With the assumptions 1 and 2 as above, and given source and channel statistics, the encoding alphabets,  the distortion functions \(\rho_{t}\) and a time horizon T, the objective is to find globally optimal encoding, decoding and memory update functions \(f^1_{1:T},f^2_{1:T},...,f^n_{1:T},g_{1:T},l^1_{1:T},l^2_{1:T},...,l^n_{1:T}\) so as to minimize
          \begin{equation} 
                    \EXP{\sum\limits_{t=1}^{T} \rho_{t}(X_{t},\hat{X_{t}})}
          \end{equation}
          For this problem we can establish, by arguments similar to those used in the problems with two encoders, the following results (Theorems 6 and 7) that are analogous to Theorem 1 and Theorem 5 respectively.
\begin{theorem}
   There exist globally optimal encoding rules of the  form  :
   \begin{equation}
   Z^i_{t}=f^i_{t}(X^i_{t},b^i_{t},\mu^i_{t}) 
   \end{equation}
   where $ b^i_t := P(A|X^i_{1:t}) $ and $ \mu^i_t := P(M^i_{t-1}|Z^i_{1:t-1},l^i_{1:t-1})$. The optimal decoding rules are of the form:
   \begin{equation}
   \hat{X}_t = \tau_t(\psi_t)
   \end{equation} 
   where $\psi_t := P(X_{t}|Y^1_{t},Y^2_t,...,Y^n_t,M^1_{t-1}, M^2_{t-1},...,M^n_{t-1})$ and \(\tau_t\) is as defined in \emph{Definition 4}. 
   \end{theorem}
   \begin{proof}
       Consider any arbitrary choice of  encoding functions for encoder 2 through encoder \(n\) and arbitrary choice of the decoding and memory update functions at the receiver. Then the problem for encoder 1 is essentially same as in the case when \(n=2\).            
\end{proof}
\begin{theorem}
   Consider Problem P3 with noiseless channel (that is, \(Y^i_t = Z^i_t\)) and perfect receiver memory (that is , \(M^i_{t-1} = Z^i_{1:t-1}\)). Then there is no loss of optimality in considering decoding rules of the form \(\hat{X}_t = \tau_t(\psi_t)\) where \(\psi_t = P(X_t|Z^1_{1:t},...,Z^n_{1:t})\) with encoders that operate as follows: \\
 For \(i=1,2,...,n\), define \(\xi^i_0 := 0\) and for \(t=1,2,...T\), 
       \begin{equation} 
        Z^i_t =f^i_t(X^i_t,b^i_t,\xi^i_{t-1})
        \end{equation}
        and
        \begin{equation} 
        \xi^i_t =\gamma^i_t(\xi^i_{t-1},Z^i_t,f^i_t(\cdot,\xi^i_{t-1}))
        \end{equation}
        where \(\gamma^i_t\) are fixed transformations (Lemma 6).
 \end{theorem} 
\begin{proof}
  The result follows from Theorem 5 using similar arguments as in the proof of Theorem 6.
\end{proof}
     
\subsection{Point-to-Point Systems}
\subsubsection {A Side Information Channel}
\begin{figure}[ht]
\begin{center}
\includegraphics[height=4cm,width=7cm]{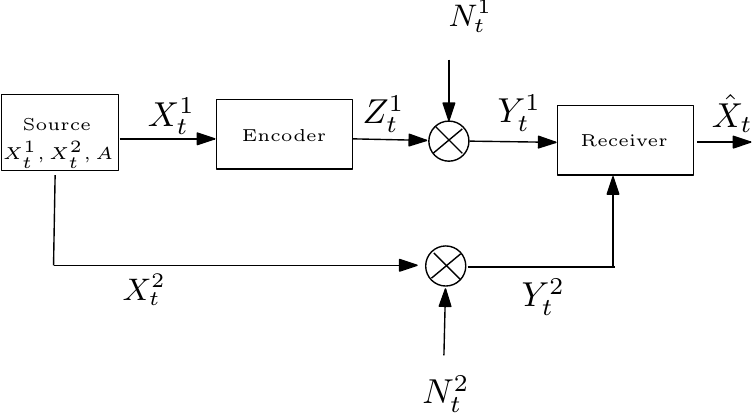}
\end{center}
\caption{Side-Information Problem} 
\label{fig:side-info}
\end{figure} 
     Consider Problem P1 or P2 with encoder~2's strategy fixed as follows:
     \[ Z^2_t = X^2_t \]
     Then the multi-terminal communication problems reduce to a point-to-point communication problems with side-information available at the receiver (See, for example, Fig.6). It is clear that the results of Theorem 1 and Theorem 2, for noisy channels, and Theorem 5, for noiseless channels, remain valid for these side-information problems as well (since they are true for any arbitrary choice of encoder~2's strategy).        
 
 \par   
\subsubsection{Unknown Transition Matrix}
       Consider a point-to-point communication system where an encoder is communicating its observations of a Markov source \(X_t\) to a receiver (Fig.7). The channel may be noisy or noiseless, the receiver may have finite memory or perfect recall. Structural results for optimal real-time encoding rules have been obtained in cases when the transition probabilities of the Markov source are known (\cite{Witsenhausen78}, \cite{Teneketzis06}). Consider now the case where the encoder observes a Markov chain \(X_t\) whose transition matrix is not known. However, the set of possible transition matrices is parameterized by a parameter \(A\) with a known prior distribution over a finite set \(\mathcal{A}\). 
The encoding functions are of the form:
       \[ Z_t = f_t(X_{1:t},Z_{1:t-1})\]
       where \(Z_t\) is the transmitted symbol at time \(t\). 
       The receiver receives a noisy version of \(Z_t\) given by
       \[Y_t=h_t(Z_t,N_t)\]
       where \(N_t\) is the noise in the channel. The receiver maintains a finite memory  that is updated as follows:
        \[ M_1 = l_1(Y_1)\] 
        \[M_t=l_t(Y_t,M_{t-1})\] 
        where \(M_t \in \mathcal{M}\), \(\forall t\).
        The receiver's estimate at time \(t\) is given as:
        \[ \hat{X}_t = g_t(Y_t,M_{t-1})\]
        A non-negative distortion function \(\rho_t(X_t,\hat{X}_t)\) measures the instantaneous distortion at time \(t\).
\begin{figure}[ht] 
\begin{center}
\includegraphics[height=2cm,width=7cm]{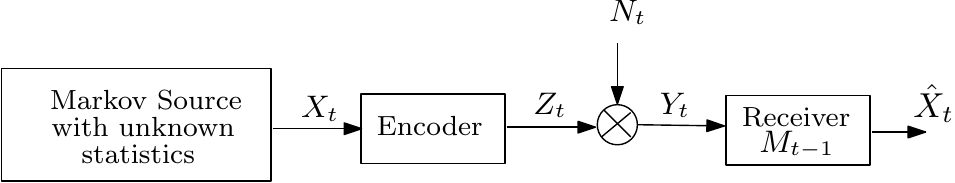}
\end{center}
\caption{Point-to-point system with unknown source statistics} 
\label{fig:p2p}
\end{figure} 
We consider the following problem:\\
\emph{Problem P4:} Given the source and receiver model as above and the noise statistics, the encoding alphabets, the channel functions \(h_t\), the distortion functions \(\rho_{t}\) and a time horizon T, the objective is to find globally optimal encoding, decoding and memory update functions \(f_{1:T},g_{1:T},l_{1:T}\) so as to minimize
          \begin{equation} 
                    J(f_{1:T},g_{1:T},l_{1:T})= \EXP{\sum\limits_{t=1}^{T} \rho_{t}(X_{t},\hat{X_{t}})}
          \end{equation}
          The methodology employed for the analysis of Problem P1 can be used to establish the following result.  
\begin{theorem}
   There exist globally optimal encoding rules of the  form  :
   \begin{equation}
   Z_{t}=f_{t}(X_{t},b_{t},\mu_{t}) 
   \end{equation}
   where $ b_t := P(A|X_{1:t}) $ and $ \mu_t := P(M_{t-1}|Z_{1:t-1},l_{1:t-1})$. The optimal decoding rules are of the form:
   \begin{equation}
   \hat{X}_t = \tau_t(\psi_t)
   \end{equation} 
   where $\psi_t := P(X_{t}|Y_{t},M_{t-1},f_{1:t},l_{1:t})$ and \(\tau_t\) is as defined in \emph{Definition 4}. 
   \end{theorem}
   \begin{proof} 
        We can view the optimization problem P4 as a special case of Problem P1 with an imaginary second encoder that makes no observations of the source and sends no message to the receiver (that is, the set \(\mathcal{X}^2 \) and \(\mathcal{Z}^2\) are empty). Thus, the results of the above theorem follow from Theorem 1 and Theorem 2.
  \end{proof}
  The methodology developed for the analysis of Problem P2 can be used to obtain the following result.
  \begin{theorem}
   Consider Problem P4 with noiseless channel (that is, \(Y_t = Z_t\)) and perfect receiver memory (that is , \(M_{t-1} = Z_{1:t-1}\)). Then there is no loss of optimality in considering encoding rules of the form:
   \[Z_t = f_t(X_t,b_t,\xi_{t-1}) \]
   where \(b_t := P(A|X_{1:t})\) and \[\xi_{t-1}:= P(X_{t-1},b_{t-1}|Z_{1:t-1})\]
  with decoding rules of the form: 
\begin{equation}
   \hat{X}_t = \tau_t(\psi_t)
   \end{equation} 
   where $\psi_t := P(X_{t}=x|Z_{1:t})$ and \(\tau_t\) is as defined in \emph{Definition 4}.
\end{theorem} 
\begin{proof}
  The result follows from Theorem 5 using similar arguments as in the proof of Theorem 8.
\end{proof}
\subsection{kth order Markov Source}
Consider Problem P1 or P2 with a source model given by the following equations:
\begin{subequations}
\begin{equation}
    X^1_{t+1} = F^1_{t}(X^1_t,X^1_{t-1},..,X^1_{t+1-k},A,W^1_t)
\end{equation}
\begin{equation}
X^2_{t+1} = F^2_{t}(X^2_t,X^2_{t-1},..,X^2_{t+1-k},A,W^2_t)
\end{equation}
\end{subequations}
Thus, conditioned on a global, time-invariant random variable \(A\), \(X^1_t\) and \(X^2_t\) are conditionally independent \(k\)th order Markov processes. It is straightforward to consider a Markovian reformulation of the source by defining 
\[ B^i_t := (X^i_1,X^i_2,...,X^i_t)\]
for \(i =1,2\) and \(t \leq k\) and  
\[ B^i_t := (X^i_t,X^i_{t-1},..,X^i_{t+1-k}) \]
for \(i=1,2\) and \(t > k\).
We then have that

\begin{equation}
    B^i_{t+1} = \tilde{F}^1_{t}(B^i_t,A,W^1_t)
\end{equation}
for \(i=1,2\). Thus, we now have a Markov system (when conditioned on \(A\)) - with \(B^i_t\) as the encoder~\(i\)'s observations - for which our structural results directly apply.      
\subsection{Communication with Finite Delay}
             Consider the models of Problem P1 or P2 with the following objective function: \[\sum\limits_{t=d+1}^{T+d} E[\rho_t(X_{t-d},\hat{X}_{t})] \] The above objective can be interpreted as the total expected distortion incurred when the receiver can allow a small finite delay, \(d\), before making its final estimate on the state of the source. Thus, the receiver produces a sequence of
               source estimates \(\hat{X}_{d+1},\hat{X}_{d+2},...,\hat{X}_{T+d}\), and incurs a distortion \(\sum\limits_{t=d+1}^{T+d} E[\rho_t(X_{t-d},\hat{X}_{t})] \).
               We can transform this problem to our problem by the following regrouping of variables.\\
               For \(i=1,2\) and \(t=1,2,..,d\) define
               \begin{equation}
                   B^i_t := (X^i_1,X^i_2,..,X^i_t)
               \end{equation}
               For \(t=d+1,...,T\), define
                  \begin{equation}
                   B^i_t := (X^i_{t-d},X^i_{t-d+1},..,X^i_t)
               \end{equation}
               and for \(t=T+1,T+2,..T+d\)
                \begin{equation}
                   B^i_t := (X^i_{t-d},X^i_{t-d+1},..,X^i_{T})
               \end{equation}
               Then, it is easily seen that conditioned on \(A\), \(B^1_t\) and \(B^2_t\) are two conditionally independent Markov chains. Moreover, the distortion function \(\rho_t(X_{t-d},\hat{X}_{t})\)
               can be expressed as \(\tilde{\rho}(B^1_t,B^2_t,A,\hat{X}_{t})\). Thus, we have modified the problem to an instance of Problem P1 or P2 with \(B^i_t\) as the encoder~\(i\)'s observation.
         
         
\section{Conclusion} \label{sec:Conc}
   We considered a real-time communication problem where two encoders make distinct partial observations of a discrete-time Markov source and communicate in real-time with a common receiver which needs to estimate some function of the state of the Markov source in real-time. We assumed a specific model for the source that arises in some applications of interest. In this model, the encoders' observations are conditionally independent Markov chains given an unobserved, time-invariant random variable. We formulated a communication problem with separate noisy channels between each  encoder and the receiver and a separated finite memory at the receiver. We obtained finite-dimensional sufficient statistics for the encoders in this problem. The structure of the source  and the receiver played a critical role in obtaining these results.
   \par
   We then considered the communication problem over noiseless channels and  perfect receiver memory. We developed a new methodology to identify structural results for this problem. The new approach highlights the importance of common information in decentralized team problems. We used the presence of common information between an encoder and the receiver to identify a sufficient statistic of the decoder that has a time-invariant domain.     
\par
We have not addressed the problem of finding globally optimal real-time encoding and decoding strategies in this paper. A sequential decomposition of the global optimization problem, for a special case of the problems formulated here, appears in \cite{Nayyar-CDC08}.    
\appendices
\section{Proof of Lemma 1}
For a realization \(x^1_{1:t}\) of \(X^1_{1:t}\), we have by definition,
\begin{align}
 b^1_t(a) = &P(A=a|x^1_{1:t}) \nonumber \\
   = &P(A=a,x^1_t|x^1_{1:t-1})/\sum\limits_{a' \in \mathcal{A}} P(A=a',x^1_t|x^1_{1:t-1}) \label{eq:lemma1eqn1} 
 \end{align}
 where we used Bayes' rule in (\ref{eq:lemma1eqn1}). The numerator in  (\ref{eq:lemma1eqn1}) can be written as,
 \begin{align}
   &P(X^1_t=x^1_t|A=a,x^1_{1:t-1}).P(A=a|x^1_{1:t-1}) \nonumber \\
   = &P(X^1_t=x^1_t|A=a,x^1_{t-1}).b^1_{t-1}(a) \label{eq:lemma1eqn2}
 \end{align}
where we used the Markov nature of \(X^1_t\) when conditioned on \(A\). Thus, for a fixed \(a\), the numerator in  (\ref{eq:lemma1eqn1}) depends only on \(x^1_t, x^1_{t-1}\) and the previous belief \(b^1_{t-1}\). Since the same factorization holds for each term in the denominator, we have that
\[b^1_{t}=\alpha^1_{t}(b^1_{t-1},X^1_{t},X^1_{t-1}), \] 
where \(\alpha^1_t, t=2,3,...,T\) are deterministic transformations.

\section{Proof of Lemma 2}
By definition of \(\mu^1_t\), we have
\begin{align}
  \mu^1_{t}(m) = P(M^1_{t-1}=m|Z^1_{1:t-1}, l^1_{1:t-1}) \nonumber \\
  = P(l^1_{t-1}(M^1_{t-2},Y^1_{t-1})=m|Z^1_{1:t-1}, l^1_{1:t-1}) \label{eq:lemma2eqn1}
\end{align}
With the memory update rules \(l^1_{1:t-1}\) fixed, the probability in (\ref{eq:lemma2eqn1}) can be evaluated from the conditional distribution $P(M^1_{t-2},Y^1_{t-1}|Z^1_{1:t-1}, l^1_{1:t-1})$. For \(m' \in \mathcal{M}^1\) and \(y \in \mathcal{Y}^1\), this conditional distribution is  given as 
\begin{align}
&P(M^1_{t-2}=m',Y^1_{t-1}=y|Z^1_{1:t-1}, l^1_{1:t-1}) \label{eq:lemma2eqn1.2} \\
= &P(Y^1_{t-1}=y|M^1_{t-2}=m',Z^1_{1:t-1}, l^1_{1:t-1}) \times \nonumber \\
 &P(M^1_{t-2}=m'|Z^1_{1:t-1}, l^1_{1:t-1}) \nonumber \\
= &P(Y^1_{t-1}=y|Z^1_{t-1}).P(M^1_{t-2}=m'|Z^1_{1:t-2}, l^1_{1:t-2}) \label{eq:lemma2eqn2} \\
= &P(Y^1_{t-1}=y|Z^1_{t-1}).\mu^1_{t-1}(m')
\end{align}
where we used the fact that the channel noise at time \(t\) (\(N^1_t\)) is independent of the past noise variables and the Markov source in (\ref{eq:lemma2eqn2}). Thus, we only need \(Z^1_{t-1}\) and \(\mu^1_{t-1}\) to form the joint belief in (\ref{eq:lemma2eqn1.2}). Consequently, we can evaluate \(\mu^1_t(m)\) just from \(Z^1_{t-1}\) and \(\mu^1_{t-1}\). Thus,
\[\mu^1_{t}=\beta^1_{t}(\mu^1_{t-1},Z^1_{t-1})\]
where \(\beta^1_t, t=2,3,...,T\) are deterministic transformations.

\section{Proof Of Lemma 5}
For fixed \(f^2_{1:T}\) and for a given realization of the received symbols \(z^1_{1:t},z^2_{1:t}
\) and the partial encoding functions \(\tilde{w}^1_{1:t}\), the receiver's belief on the state of the source at time \(t\) is given as:
\begin{equation}
  \psi_{t}(x) := P(X_{t}=x|z^{1}_{1:t},z^{2}_{1:t},\tilde{w}^1_{1:t},f^{2}_{1:t})
\end{equation}
where \(x=(x^1,x^2,a)\).
Using Bayes' rule, we have
\begin{multline}\label{eq:P2lemma3}
\psi_{t}(x) = P(X_{t}=x,Z^{2}_{1:t}=z^2_{1:t}|z^{1}_{1:t},\tilde{w}^1_{1:t},f^{2}_{1:t})/\\
\sum\limits_{x' \in \mathcal{X}}P(X_t=x',Z^{2}_{1:t}=z^2_{1:t}|z^{1}_{1:t},\tilde{w}^1_{1:t},f^{2}_{1:t})
\end{multline}

The numerator in right hand side of (\ref{eq:P2lemma3}) can be written as
\begin{align}\label{eq:six}
&P(Z^{2}_{1:t}=z^2_{1:t}|z^{1}_{1:t},X_{t}=x,\tilde{w}^1_{1:t},f^{2}_{1:t})  \nonumber \\
&\times P(X_{t}=x|z^{1}_{1:t},\tilde{w}^1_{1:t},f^{2}_{1:t}) \nonumber \\
 = &P(Z^{2}_{1:t}=z^2_{1:t}|X^2_t=x^2,A=a,f^{2}_{1:t}) \nonumber \\ &\times P(X^2_{t}=x^2|A=a)  
 \times P(X^1_t=x^1_t,A=a|z^{1}_{1:t},\tilde{w}^1_{1:t})
\end{align}
where we used conditional independence of \(Z^{2}_{1:t},X^2_t\) and \(Z^{1}_{1:t},X^1_{t}\) given \(A\) for the first term in (\ref{eq:six}) and the fact that \(X_t = (X^1_t,X^2_t,A)\) in the second term of left hand side of (\ref{eq:six}).

Since the second encoder is fixed, the first term in the right hand side of (\ref{eq:six}) is a known statistic which depends on \(z^2_{1:t}\). The second term is again a known source statistic. Consider the last term in (\ref{eq:six}). It can be expressed as follows:
\begin{align}
 &\int\limits_{b' \in \Delta(\mathcal{A})} P(X^1_t=x^1_t,A=a,b^1_t=b'|z^{1}_{1:t},\tilde{w}^1_{1:t}) \nonumber \\
 = &\int\limits_{b' \in \Delta(\mathcal{A})} [P(A=a|b^1_t=b',x^1_t,z^{1}_{1:t},\tilde{w}^1_{1:t}) \nonumber \\
 & \times P(X^1_t=x^1_t,b^1_t=b'|z^{1}_{1:t},\tilde{w}^1_{1:t})] \\
 = &\int\limits_{b' \in \Delta(\mathcal{A})}b'(a) \times P(X^1_t=x^1_t,b^1_t=b'|z^{1}_{1:t},\tilde{w}^1_{1:t}) \nonumber \\
 = &\int\limits_{b' \in \Delta(\mathcal{A})}b'(a) \times \tilde{\xi}^1_t(x^1_t,b')\label{eq:P2lemma3.01}
\end{align}
Similar representations also hold for each term in the denominator of (\ref{eq:P2lemma3}).
It follows then that with a fixed \(f^2_{1:t}\), \(\psi_t(x)\) depends only on the realization of second encoder's messages \(Z^2_{1:t}\) and \(\xi^1_t\). Thus, from (\ref{eq:six}) and (\ref{eq:P2lemma3.01}), we conclude that \(\psi_{t}\) can be evaluated from \(\xi^1_t\) and \(Z^2_{1:t}\) by means of deterministic transformations. We will call this overall transformation as \(\delta_{t}\). Thus, we have
\begin{equation}
\psi_t=\delta_t(\xi^1_t,Z^2_{1:t})
\end{equation}
 Since the estimate \(\hat{X}_t\) is a function of \(\psi_t\) (cf. Theorem 3), we conclude that
\[\hat{X}_t = \tau_t(\delta_t(\xi^1_t,Z^2_{1:t}))\]

\section{Proof of Lemma 6}
1) Consider a realization \(z^1_{1:t}\) and \(\tilde{w}^1_{1:t}\).\\
By definition, the realization of \(\xi^1_{t-1}\) is given as
\begin{equation}
 \tilde{\xi}^1_t(x^1_t,\tilde{b}^1_t) = P(X^1_t=x^1_t,b^1_t=\tilde{b}^1_t|z^1_{1:t},\tilde{w}^1_{1:t})
\end{equation}
Using Bayes' rule, we have
\begin{align}
&\tilde{\xi}^1_t(x^1_t,\tilde{b}^1_t) = P(X^1_t=x^1_t,b^1_t=\tilde{b}^1_t,Z^1_t=z^1_t|z^1_{1:t-1},\tilde{w}^1_{1:t}) \nonumber\\
 &/ \sum\limits_{\scriptscriptstyle x' \in \mathcal{X}^1}\int\limits_{\substack{\scriptscriptstyle b' \in \Delta(\mathcal{A})}}P(X^1_t=x',b^1_t=b',Z^1_t=z^1_t|z^1_{1:t-1},\tilde{w}^1_{1:t}) \label{eq:neweq1}
\end{align}
We can write the numerator as:
\begin{align}
&P(Z^1_t=z^1_t|X^1_t=x^1_t,b^1_t=\tilde{b}^1_t,z^1_{1:t-1},\tilde{w}^1_{1:t}) \nonumber \\
 &\times P(X^1_t=x^1_t,b^1_t=\tilde{b}^1_t|z^1_{1:t-1},\tilde{w}^1_{1:t}) \nonumber \\
&=  P(Z^1_t=z^1_t|X^1_t=x^1_t,b^1_t=\tilde{b}^1_t,\tilde{w}^1_{t}) \nonumber \\
&\times P(X^1_t=x^1_t,b^1_t=\tilde{b}^1_t|z^1_{1:t-1},\tilde{w}^1_{1:t}) \label{eq:P2lemma1.0}
\end{align}
the first term in (\ref{eq:P2lemma1.0}) is true since \(z^1_t=w^1_t(x^1_t,\tilde{b}^1_t)\). The second term in (\ref{eq:P2lemma1.0}) can be further written as:
\begin{align}
\sum\limits_{\substack{\scriptscriptstyle x'' \in \mathcal{X}^1,\\ \scriptscriptstyle a \in \mathcal{A}}}\int\limits_{\scriptscriptstyle b'\in \Delta(\mathcal{A})} &P(X^1_t=x^1_t,b^1_t=\tilde{b}^1_t, X^1_{t-1}=x'' \nonumber \\
& ,A=a,b^1_{t-1}=b'|z^1_{1:t-1},\tilde{w}^1_{1:t}) \nonumber \\
= \sum\limits_{\substack{\scriptscriptstyle x'' \in \mathcal{X}^1,\\ \scriptscriptstyle a \in \mathcal{A}}}\int\limits_{\scriptscriptstyle b'\in \Delta(\mathcal{A})}
&[P(b^1_t=\tilde{b}^1_t|b^1_{t-1}=b',X^1_{t}=x^1_t,X^1_{t-1}=x'') \nonumber \\
& \times P(X^1_t=x^1_t|A=a, X^1_{t-1}=x'') \nonumber \\
& \times P(A=a|b^1_{t-1}=b',X^1_{t-1}=x'', \nonumber \\ &z^1_{1:t-1},\tilde{w}^1_{1:t-1}) \nonumber \\ &\times P(X^1_{t-1}=x'', b^1_{t-1}=b'|z^1_{1:t-1},\tilde{w}^1_{1:t-1})] \label{eq:new_app1}\\
= \sum\limits_{\substack{\scriptscriptstyle x'' \in \mathcal{X}^1,\\ \scriptscriptstyle a \in \mathcal{A}}}\int\limits_{\scriptscriptstyle b'\in \Delta(\mathcal{A})} &[P(b^1_t=\tilde{b}^1_t|b^1_{t-1}=b',X^1_{t}=x^1_t,X^1_{t-1}=x'') \nonumber \\
& \times P(X^1_t=x^1_t|A=a, X^1_{t-1}=x'') \nonumber \\
& \times P(A=a|b^1_{t-1}=b') \times \tilde{\xi}^1_{t-1}(x'',b')] \label{eq:P2lemma1.1}
\end{align}
where we used Lemma 1 and the Markov property of \(X^1_t\) given \(A\) in (\ref{eq:new_app1}). The first term in (\ref{eq:P2lemma1.1}) is simply 1 or 0 since \(b^1_t\) is a deterministic function of \(b^1_{t-1}\), \(X^1_t\) and \(X^1_{t-1}\). The second term is a known source statistic and the third term is \(b'(a)\). Similar expressions hold for the denominator in (\ref{eq:neweq1}). Thus, from (\ref{eq:neweq1})-(\ref{eq:P2lemma1.1}), we conclude that to evaluate \(\xi^1_t(x^1_t,\tilde{b}^1_t)\) we only need \(Z^1_t,w^1_t\) and \(\xi^1_{t-1}\).  This establishes equation (\ref{eq:P2lemma6a}).\\
\par
2) With encoder~2's strategy fixed, the expected instantaneous cost from the coordinator's perspective is given as:
\begin{align}
&\EXP{\rho_{t}(X_{t},\hat{X}_{t})|z^1_{1:t},\tilde{w}^1_{1:t}} \nonumber \\
= &\EXP{\rho_{t}(X^1_{t},X^2_t,A,\tau_{t}(\delta_{t}(\xi^1_t,Z^2_{1:t})))|z^1_{1:t},\tilde{w}^1_{1:t},\tilde{\xi}^1_t}, \label{eq:P2lemma3.0} 
\end{align}
since \(\tilde{\xi}^1_t\) is a function of \(z^1_{1:t},\tilde{w}^1_{1:t}\), hence it can be included in the conditioning variables. Thus, the only random variables in the above expectation are \(X^1_t,X^2_t,A\) and \(Z^2_{1:t}\). Therefore, the above expectation is a function of the following probability mass function:
\begin{align}
 &P(X^1_t=x^1_t,X^2_t=x^2_t,A=a,Z^2_{1:t}=z^2_{1:t}|z^1_{1:t},\tilde{w}^1_{1:t},\tilde{\xi}^1_t) \nonumber \\
= &P(Z^2_{1:t}=z^2_{1:t},X^2_t=x^2_t|A=a,X^1_t=x^1_t,z^1_{1:t},\tilde{w}^1_{1:t},\tilde{\xi}^1_t) \nonumber \\
&\times P(X^1_t=x^1_t,A=a|z^1_{1:t},\tilde{w}^1_{1:t},\tilde{\xi}^1_t) \nonumber \\
= &P(Z^2_{1:t}=z^2_{1:t},X^2_t=x^2_t|A=a) \times  \nonumber \\
&\int\limits_{\scriptscriptstyle b' \in \Delta(\mathcal{A})} [ P(X^1_t=x^1_t,A=a,b^1_t=b'|z^{1}_{1:t},\tilde{w}^1_{1:t},\tilde{\xi}^1_t)] \nonumber \\
 = &P(Z^2_{1:t}=z^2_{1:t},X^2_t=x^2_t|A=a) \times \nonumber \\
&\int\limits_{b' \in \Delta(\mathcal{A})} [P(A=a|b^1_t=b',z^{1}_{1:t},\tilde{w}^1_{1:t},\tilde{\xi}^1_t) \label{eq:P2lemma3.1} \\
 & \times P(X^1_t=x^1_t,b^1_t=b'|z^{1}_{1:t},\tilde{w}^1_{1:t},\tilde{\xi}^1_t)] \nonumber
 \end{align}
\begin{align}
 = &P(Z^2_{1:t}=z^2_{1:t},X^2_t=x^2_t|A=a) \times \int\limits_{b' \in \Delta(\mathcal{A})}[b'(a)  \nonumber \\
 &\times P(X^1_t=x^1_t,b^1_t=b'|z^{1}_{1:t},\tilde{w}^1_{1:t},\tilde{\xi}^1_t)] \nonumber \\
 = &P(Z^2_{1:t}=z^2_{1:t},X^2_t=x^2_t|A=a) \times \int\limits_{b' \in \Delta(\mathcal{A})}[b'(a) \nonumber \\
 &\times \tilde{\xi}^1_t(x^1_t,b')] \label{eq:P2lemma3.2}
\end{align}
where we used conditional independence of the encoder's observations and actions given \(A\) in (\ref{eq:P2lemma3.1}).
In (\ref{eq:P2lemma3.2}), the first term is a fixed statistic when encoder~2's strategy is fixed and the second term depends only on \(\tilde{\xi}^1_t\). Thus, the expectation in (\ref{eq:P2lemma3.0}) can be evaluated using \(\tilde{\xi}^1_t\). This establishes the second part of the Lemma (equation \ref{eq:P2lemma6b}).
\section{ Proof of Theorem 4}
For \(\xi \in \Delta(\mathcal{X}^1 \times \Delta(\mathcal{A}))\), define the following functions:
\begin{align}
V_{T}(\xi) = &\overline{\rho}_{T}(\xi) \label{eq:dp1} \\
V_{t-1}(\xi) = &\overline{\rho}_{t-1}(\xi) +  \nonumber \\
&\inf_{w}[ \EXP{ V_t(\gamma_t(\xi,Z^1_t,w^1_t ))|\xi^1_{t-1}=\xi, w^1_t=w}] \label{eq:dp2}
\end{align}
for \(t=T,T-1, \ldots, 2\) and
\begin{equation}\label{eq:v0}
V_0 = \inf_{w}[ \EXP{ V_1(\gamma_t(\xi,Z^1_1,w^1_1 ))|w^1_1=w}]
\end{equation} The functions \(\overline{\rho}_{t}\) and \(\gamma_t\) are from Lemma 6. Note that the infimum in (\ref{eq:dp2})  is over all functions from the space \((\mathcal{X}^1 \times \Delta(\mathcal{A}))\) to the space \(\mathcal{Z}^1\) and the infimum in (\ref{eq:v0}) is over all functions from the space \(\mathcal{X}^1\) to the space \(\mathcal{Z}^1\).
\par
Consider an arbitrary selection rule \(\Lambda := (\Lambda_t, t=1,2...,T)\) for the coordinator. That is, the coordinator selects the partial-encoding function at time \(t\) as follows:
\begin{equation}
w^1_t = \Lambda_t(Z^1_{1:t-1},w^1_{1:t-1})
\end{equation}

Then the coordinator's expected cost to go from time \(t\) onwards under the selection rule \(\Lambda\) is given as :
\begin{equation}
J_t(Z^1_{1:t},w^1_{1:t}) = \EXP{\sum\limits_{k=t}^{T}\rho_{k}(X_{k},\hat{X}_{k})|Z^1_{1:t},w^1_{1:t}}
\end{equation}
for \(t=1,2, \ldots, T\). Also, the overall expected cost under selection rule \(\Lambda\) is
\begin{equation}
J_0 = \EXP{J_1(Z^1_1,w^1_1)}
\end{equation}

We will show for all \(t=T,T-1,..\ldots, 1\), we have the following inequality
\begin{equation}
V_t(\xi^1_{t}) \leq J_t(Z^1_{1:t},w^1_{1:t}) \label{eq:dplemma}
\end{equation}
where the \(\xi^1_t\) is the belief on \(X^1_t,b^1_t\) conditioned on \(Z^1_{1:t},w^1_{1:t}\).
We proceed by backward induction. At time \(T\), we have
\begin{align}
J_T(Z^1_{1:T},w^1_{1:T}) = &\EXP{\rho_{T}(X_{T},\hat{X}_{T})|Z^1_{1:T},w^1_{1:T}} \\
 = &\overline{\rho}_{T}(\xi^1_T) = V_T(\xi^1_T) \label{eq:dpproof1}
\end{align}
where we used part 2 of Lemma 6 (equation \ref{eq:P2lemma6b}) in (\ref{eq:dpproof1}).  Thus (\ref{eq:dplemma}) is true for \(t=T\).
Assume that (\ref{eq:dplemma}) is true for time \(t\). At \(t-1\), for a realization \(z^1_{1:t-1},\tilde{w}^1_{1:t-1}\), we have
\begin{align}
\lefteqn{J_{t-1}(z^1_{1:t-1},\tilde{w}^1_{1:t-1})} \nonumber \\
= &\EXP{\sum\limits_{k=t-1}^{T}\rho_{k}(X_{k},\hat{X}_{k})|z^1_{1:t-1},\tilde{w}^1_{1:t-1}} \\
= &\EXP{\rho_{t-1}(X_{t-1},\hat{X}_{t-1})|z^1_{1:t-1},\tilde{w}^1_{1:t-1}} + \nonumber \\ &\EXP{\sum\limits_{k=t}^{T}\rho_{k}(X_{k},\hat{X}_{k})|z^1_{1:t-1},\tilde{w}^1_{1:t-1}}  \\
= &\overline{\rho}_{t-1}(\tilde{\xi}^1_{t-1}) +  \nonumber \\ &\EXP{\EXP{\sum\limits_{k=t}^{T}\rho_{k}(X_{k},\hat{X}_{k})|Z^1_{1:t},w^1_{1:t}}|z^1_{1:t-1},\tilde{w}^1_{1:t-1}} \label{eq:dpproof2} \\
\geq &\overline{\rho}_{t-1}(\xi^1_{t-1}) + \EXP{V_t(\xi^1_t)|z^1_{1:t-1},\tilde{w}^1_{1:t-1}} \label{eq:dpproof3}
\end{align}
where we used part 2 of lemma 6 for the first term in (\ref{eq:dpproof2}) and the induction hypothesis at time \(t\) for the second term in (\ref{eq:dpproof3}). We will focus on the second term in (\ref{eq:dpproof3}). We have
\begin{align}
&\EXP{V_t(\xi^1_t)|z^1_{1:t-1},\tilde{w}^1_{1:t-1}} \nonumber \\
&=\EXP{V_t(\xi^1_t)|z^1_{1:t-1},\tilde{w}^1_{1:t-1},\tilde{\xi}^1_{t-1},\tilde{w}^1_t} \label{eq:temp1}
\end{align}
Note that we have included \(\tilde{\xi}^1_{t-1}\) and \(\tilde{w}^1_t\) in the conditioning of the right hand side of (\ref{eq:temp1}) since under the selection rule \(\Lambda\), they are functions of the original conditioning terms \(z^1_{1:t-1},\tilde{w}^1_{1:t-1}\). Further, using Lemma 6 for \(\xi^1_t\) in (\ref{eq:temp1}), we get
\begin{align}
&\EXP{V_t(\xi^1_t)|z^1_{1:t-1},\tilde{w}^1_{1:t-1},\tilde{\xi}^1_{t-1},\tilde{w}^1_t} \nonumber \\
& = \EXP{V_t(\gamma_{t}(\tilde{\xi}^1_{t-1}, Z^1_t, w^1_t)|z^1_{1:t-1},\tilde{w}^1_{1:t-1},\tilde{\xi}^1_{t-1},\tilde{w}^1_t} \nonumber \\
&\geq \inf_w  \EXP{ V_t(\gamma_t(\tilde{\xi}^1_{t-1},Z^1_t,w ))|z^1_{1:t-1},\tilde{w}^1_{1:t-1},\tilde{\xi}^1_{t-1}, w^1_t=w} \nonumber \\
&= \inf_w \mathds{E} \{V_t(\gamma_{t}(\tilde{\xi}^1_{t-1}, w(X^1_t,b^1_t), w))|z^1_{1:t-1}, \nonumber \\
  & \tilde{w}^1_{1:t-1},\tilde{\xi}^1_{t-1},w^1_t=w\} \label{eq:dpproof4}
\end{align}
We will now show that the right hand side in (\ref{eq:dpproof4}) is same as the second term in (\ref{eq:dp2}) evaluated at \(\tilde{\xi}^1_{t-1}\). 
. The expectation in (\ref{eq:dpproof4}) depends only on \(P(X^1_t,b^1_t|z^1_{1:t-1},\tilde{w}^1_{1:t-1},\tilde{\xi}^1_{t-1},w^1_t=w)\). Now,
\begin{align}
&P(X^1_t,b^1_t|z^1_{1:t-1},\tilde{w}^1_{1:t-1},\tilde{\xi}^1_{t-1},w^1_t=w) \nonumber \\
= \sum\limits_{\substack{\scriptscriptstyle x' \in \mathcal{X}^1,\\ \scriptscriptstyle a \in \mathcal{A}}} \int\limits_{\scriptscriptstyle b' \in \Delta(\mathcal{A})} [&P(X^1_t,b^1_t, X^1_{t-1}=x',A=a,b^1_{t-1}=b'|z^1_{1:t-1}, \nonumber\\  &\tilde{w}^1_{1:t-1},\tilde{\xi}^1_{t-1},w^1_t=w)] \nonumber \\
= \sum\limits_{\substack{\scriptscriptstyle x' \in \mathcal{X}^1,\\ \scriptscriptstyle a \in \mathcal{A}}} \int\limits_{\scriptscriptstyle b' \in \Delta(\mathcal{A})} &[P(b^1_t|X^1_t,b^1_{t-1}=b',X^1_{t-1}=x')  \nonumber \\
&\times P(X^1_t|X^1_{t-1},A=a) \times P(A=a|b^1_{t-1}=b') \nonumber \\ \times \tilde{\xi}^1_{t-1}(x',b')] \label{eq:june2.1}
\end{align}
where we used the Lemma 1 and Markov nature of \(X^1_t\) when conditioned on \(A\) in right hand side of (\ref{eq:june2.1}). The right hand side of (\ref{eq:june2.1}) depends only on \(\tilde{\xi}^1_{t-1}\) (and the known source statistics). Thus, the probability \(P(X^1_t,b^1_t|z^1_{1:t-1},\tilde{w}^1_{1:t-1},\tilde{\xi}^1_{t-1},w^1_t=w)\) is the same as the probability \(P(X^1_t,b^1_t|\tilde{\xi}^1_{t-1},w^1_t=w)\), hence the expression in the right hand side of (\ref{eq:dpproof4}) is the same as
\begin{equation} \nonumber 
  \inf_w[ \mathds{E} \{V_t(\gamma_{t}(\tilde{\xi}^1_{t-1}, w(X^1_t,b^1_t), w))|\tilde{\xi}^1_{t-1},w^1_t=w\}]
\end{equation}
 which is the second term in (\ref{eq:dp2}) evaluated at \(\tilde{\xi}^1_{t-1}\). Therefore, using (\ref{eq:dpproof3}), we get
\begin{align}
\lefteqn{J_{t-1}(Z^1_{1:t-1},w^1_{1:t-1})} \nonumber \\
 \geq &\overline{\rho}_{t-1}(\xi^1_{t-1}) +  \nonumber \\
&\inf_{w}[ \EXP{ V_t(\gamma_t(\xi^1_{t-1},Z^1_t,w^1_t ))|\xi^1_{t-1}, w^1_t=w}] \nonumber \\
 &= V_{t-1}(\xi^1_{t-1})
 \end{align}
 This completes the induction argument. Thus, we have that under any selection rule for the coordinator
 \begin{align}
 V_1(\xi^1_{1}) \leq J_1(Z^1_{1},w^1_{1}) \label{eq:june2.2}
 \end{align}
 Taking expectations on both side of (\ref{eq:june2.2}) and using the definition of \(V_0\), we get
 \[ V_0 \leq J_0 \]
 for any selection rule \(\Lambda\) for the coordinator. Now a selection rule found using equations (\ref{eq:dp1}) and (\ref{eq:dp2}) that at each step selects a \(w^1_t\) based on \(\xi^1_{t-1}\) such that it is at least as close to the infimum \(V_t\) as \(J_t\) will achieve a performance that is no worse than \(\Lambda\). This establishes the Theorem.

\section*{Acknowledgments}
  This research was supported in part by NSF Grant CCR-0325571 and NASA Grant NNX06AD47G.
\nocite{Witsenhausen73}
\nocite{Aditya}
\nocite{Nayyar-CDC08}
\nocite{Gaarder}
\bibliographystyle{IEEEtran}
\bibliography{myref}
\end{document}